\documentclass{article}

\usepackage{arxiv}

\usepackage{graphicx}
\usepackage[utf8]{inputenc}
\usepackage{amsmath}
\usepackage[section]{placeins}
\usepackage{array}
\usepackage{booktabs}
\usepackage{xcolor}
\usepackage{units}
\usepackage{amsfonts}
\usepackage{mathtools}
\usepackage{xifthen}
\usepackage{listings}
\usepackage{lineno}
\usepackage[colorlinks=true]{hyperref}
\usepackage{longtable}
\modulolinenumbers[5]
\usepackage{setspace}
\usepackage{changepage}
\usepackage{makecell}
\usepackage{multirow}
\usepackage{tikz}
\usepackage{url}

\title{A Generic Methodology for the Statistically Uniform \& Comparable Evaluation of Automated Trading Platform Components}

\author{
 Artur Sokolovsky \\
  School of Computing\\
  Newcastle University\\
  1 Science Square, NE4 5TG, UK \\
  \texttt{artur.sokolovky@gmail.com} \\
   \And
 Luca Arnaboldi \\
  School of Informatics\\
  University of Edinburgh\\
  10 Crichton St, EH8 9AB, UK \\
  \texttt{luca.arnaboldi@ed.ac.uk} \\
}

\begin{document}
\maketitle
\begin{abstract}
\textbf{Introduction:} Although machine learning approaches have been widely used in the field of finance, to very successful degrees, these approaches remain bespoke to specific investigations and opaque in terms of explainability, comparability, and reproducibility. 
    
\textbf{Objectives:} The primary objective of this research was to shed light upon this field by providing a generic methodology that was investigation-agnostic and interpretable to a financial markets practitioner, thus enhancing their efficiency, reducing barriers to entry, and increasing the reproducibility of experiments. 
The proposed methodology is showcased on two automated trading platform components. Namely, price levels, a well-known trading pattern, and a novel 2-step feature extraction method. 

\textbf{Methods:} This proposed a generic methodology, usable across markets, the methodology relies on hypothesis testing, which is widely applied in other social and scientific disciplines to effectively evaluate the concrete results beyond simple classification accuracy.
The first hypothesis was formulated to evaluate whether the selected trading pattern is suitable for use in the machine learning setting.
The second hypothesis allows to systematically assess whether the proposed feature extraction method leads to any statistically significant improvement in the automated trading platform performance. 

\textbf{Results:} Experiments were conducted across, 10 contracts, 3 feature spaces, and 3 rebound configurations (for feature extraction), resulting in 90 experiments. 
Across the experiments we found  that the use of the considered trading pattern in the machine learning setting is only partially supported by statistics, resulting in insignificant effect sizes (Rebound 7 - $0.64 \pm 1.02$, Rebound 11 $0.38 \pm 0.98$, and rebound 15 - $1.05\pm 1.16$), but allowed the rejection of the null hypothesis based on the outcome of the statistical test. 
While the results of the proposed 2-step feature extraction looked promising at first sight, statistics did not support this, this demonstrated the usefulness of the proposed methodology. 
Additionally, we obtained SHAP values for the considered models, providing insights for adjustments to the feature space.

\textbf{Conclusion:} We showcased the generic methodology on a US futures market instrument and provided evidence that with this methodology we could easily obtain informative metrics beyond the more traditional performance and profitability metrics. The interpretability of these results allows the practitioner to construct more effective automated trading pipelines by analysing their strategies using an intuitive and statistically sound methodology. This work is one of the first in applying this rigorous statistically-backed approach to the field of financial markets and we hope this may be a springboard for more research. A full reproducibility package is shared.
\end{abstract}


\section{Introduction}

As machine learning (ML) changes and takes over virtually every aspect of our lives, we are now able to automate tasks that previously were only possible with human intervention.
A field in which it has quickly gained traction and popularity is finance~\cite{easley2019microstructure}.

Most current-day trading is done electronically, through various available applications~\cite{cliff2011technology}.
Market data is propagated by the trading exchanges and handled by specialised trading feeds to keep track of trades, bids and asks by the participants of the exchange.
Different exchanges provide data in varying formats with different levels of granularity (e.g. order book) and can be accessed live through APIs or downloaded for historical data.

\begin{figure}[htbp!]
    \centering
    \includegraphics[width=0.75\textwidth]{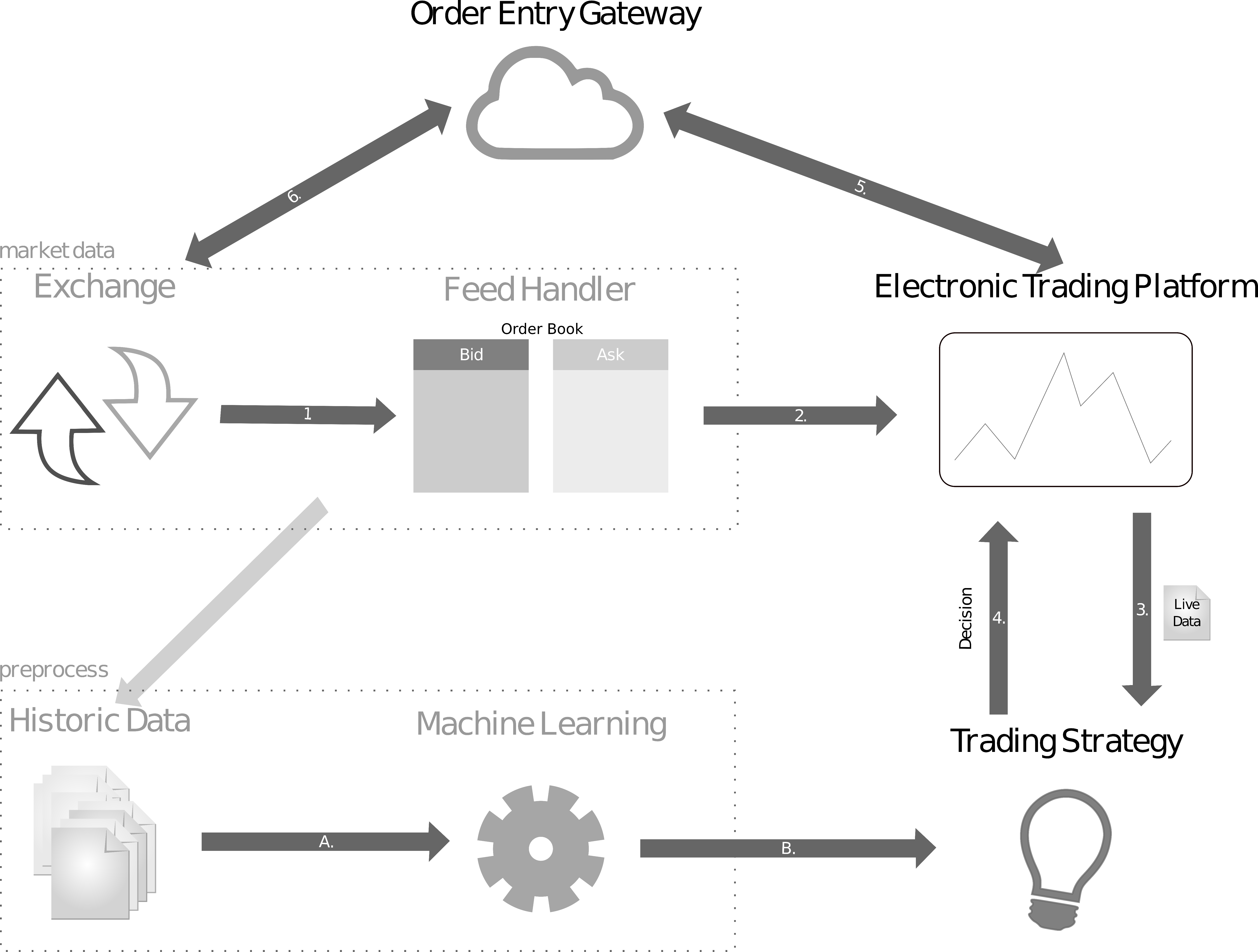}
    \caption{Schematic representation of the Automated Trading Platform components}
    \label{fig:atp-overview}
\end{figure}

A diagram of what an Automated Trading Platform (ATP) looks like in practice is presented in Fig~\ref{fig:atp-overview}.

\paragraph{Research gap} Machine learning in finance has been largely dominated by deep learning approaches~\cite{sezer2020financial,ozbayoglu2020deep}. 
The consensus has been that deep learning has continuously been the best performing predictor class, not only across financial time series but across a myriad of other disciples~\cite{sezer2020financial}. 
However, it is well known that deep learning techniques are at least difficult to explain and provide no intuition in regards to their decisions~\cite{adadi2018peeking}. 
As an outcome of this, there has been a spike in research into the explainability and accountability of machine learning~\cite{raji2020closing,ahmad2020fairness}.
Whilst the model performance is no doubt an important factor in designing profitable patterns, the inherent trade-off with explainability cannot be dismissed lightly. 
Following the decisions of a deep learning model is non-trivial, and even deemed unfeasible in most scenarios~\cite{biran2017human}.
When high financial risks are in place, it is important to be able to understand the root cause of a potential loss and adapt the strategy accordingly, however, this is incredibly difficult if not impossible using deep learning approaches~\cite{wieringa2020account,adadi2018peeking}.
What is instead desirable is to be able to measure the effectiveness of the selected trading pattern, the feature space and the model predictions, thus allowing the full power of machine learning to be supported by expert insights. 
A fine balance is to use modern, highly accurate approaches, such as tree boosting~\cite{prokhorenkova2018catboost}, but that can also be interpreted more effectively through post-hoc explanation methods~\cite{adadi2018peeking}.
An example of a technique relying on expert patterns to trade using machine learning approaches~\cite{leigh2002analysis} shows that these can be used as a direct indicator of market performance, providing profitable returns and concrete evidence of success, this is a technique that is much more approachable to the average ML practitioner.

To assess these approaches and apply the designed components of APT successfully, statistical techniques should be applied to provide the desired interpretability and comparability.
Statistical analysis such as this, on the data and predictors, is often applied to scientific fields such as medicine and psychology~\cite{cooper2018reporting}, but is much rarer in machine learning contexts.

As one can see further in the text, the proposed approach allows for the careful selection of trading criteria, greatly simplifying the analysis process, improving the uniformity of experiments, allowing by-design comparability, and justifying every layer of added complexity.

\paragraph{Study aim}
    The aim of the study is comprised of multiple aspects: i) introduce and showcase a methodology for reproducible and comparable evaluation of the ATP components, ii) automate a well-known trading pattern extraction method (price levels~\cite{pruitt1988crisma, ferris1988predicting}) and evaluate it following the proposed methodology, iii) propose and evaluate a 2-step way of feature design in the context of the automated trading system.
    
\paragraph{Research Question 1}
    As the first research question, we aim to assess whether the considered trading pattern can be classified with performance better than the baseline, always-positive class model. \textbf{RQ1:}
    \textit{Is it feasible to classify the extracted price extrema with a precision better than the baseline?}
   
    Answering the research question allows us to tell whether the proposed trading pattern is suitable for the machine learning setting. Alternatively, whether the added complexity of introducing machine learning methods in the pipeline can be justified. 
    There are a few scenarios possible. For instance, it might be the case that the proposed trading pattern in itself leads to a good baseline performance, which, however, cannot be further improved with machine learning methods. At this point, one concludes that the use of machine learning is not justified. 
    Alternatively, the pattern might perform poorly on its own but allows great classification performance improvement using machine learning tools. In such a case, the added complexity is well-justified.
    Finally, the worst case is that the pattern performs poorly and cannot be further improved using machine learning methods.
    We believe that answering this research question is an essential initial step for trading pattern design, as without this information any following analysis is lacking the necessary context.
    
\paragraph{Research Question 2}
    As the second research question, we consider a domain knowledge-based 2-step extraction method that is believed to capture a more complete representation of the market and investigate whether it gives any benefit in comparison to using any of the extraction steps alone.
    \textbf{RQ2:}\textit{ Does the use of the 2-step feature extraction improve the price level classification performance with respect to the individual steps? }

\paragraph{Outline} The rest of the paper is structured as follows: Section 2) provides the financial and machine learning background needed to understand our approach;
Section 3) provides descriptions of some related work; 
Section 4) presents the studied data and the study methodology for the automated market profiling approach and assessment methodology; 
Section 5) details our constructed APT results; 
in Section 6) a discussion on results, their implications and limitations is provided; 
and Section 7) concludes the work.

\paragraph{Contributions} The contributions of this work are the following:
\begin{enumerate}
    \item  We propose a methodology to design and statistically evaluate components of an automated trading platform using machine learning;
    \item We present an automated trading pattern extraction method with their further classification into trading scenarios;
    \item We propose and evaluate the performance of a domain knowledge-based feature space design method using the proposed methodology.
\end{enumerate}

\section{Background and related works}
\label{sec:background}

In this work, we build upon and make use of several different concepts from the fields of finance, machine learning and statistical reproducibility.
In this section, we include background details required to understand each of these topics as well as an overview of the relevant literature in these areas.
We finally conclude the section by discussing similar methodologies and how they differ from our proposed approach.

\subsection{Financial time series analysis and trading}
 
This section breaks down how data is handled, processed and used in the context of financial analysis.
This includes the pre-processing phase pattern extraction, trading, evaluation methods and features.
    
\subsubsection{Data pre-processing}
    
To prepare data for processing, the raw data is structured into predetermined formats to make it easier for machine learning algorithms to digest.
There are several ways to group data, and various features may be aggregated.
The main idea is to identify a window of interest based on some heuristic, and then aggregate the features of that window to get a representation, called \textit{Bar}.
Bars may contain several features and it is up to the individual to decide what features to select, common features include \textit{Bar start time}, \textit{Bar end time}, \textit{Sum of Volume}, \textit{Open Price}, \textit{Close Price}, \textit{Min} and \textit{Max} (usually called High and Low) prices, and any other features that might help characterise the trading performed within this window~\cite{de2018advances}.
How the bars are selected has a distinct impact on performance e.g. using time as a metric for the bar window is affected by \textit{active} and \textit{non-active} trading periods, so you might find that only some bars are useful using this methodology.

\subsubsection{Discovery of the price levels}
\label{sec:background:PriceExtrema}

In mathematics, an extremum is any point at which the value of a function is largest (a maximum) or smallest (a minimum).
These can either be local or global extrema.
At the local extremum, the value is larger/lower at immediately adjacent points, while at a global extremum the value of the function is larger than its value at any other point in the interval of interest.
If one wants to maximise their profits theoretically, they would intend to identify an extremum (or a price level) and trade at that point of optimality i.e. the peak. 
This is one of the many ways of defining the points of optimality.

As far as the algorithms for an ATP are considered, they often work in the trading context, so finding a global extremum serves little purpose. 
Consequently, local extrema within a pre-selected window are instead chosen.
Several complex algorithms exist for this with use cases in many fields such as biology~\cite{du2006improved}.
However, the objective is quite simple: identify a sample for which neighbours on each side have lower values for maxima and higher values for minima. 
This approach is very straightforward and can be implemented with a linear search.
In the case where there are flat peaks, which means several entries are of equal value the middle entry may be selected.
Two further metrics of interest are, the prominence and width of a peak.
The prominence of a peak measures how much a peak stands out from the surrounding baseline of the near entries, and is defined as the vertical distance between the peak and lowest point.
The width of the peak is the distance between each of the lower bounds of the peak, signifying the duration of the peaks.
In the case of peak classification, these measures can aid a machine learning estimator to relate the obtained features with the discovered peaks, this avoids attempts to directly relate properties of narrow or less prominent peaks with wider or more prominent peaks.
These measures provide a context in terms of price level characteristics. One might expect that they would aid the classification of trading points.

We are by no means the first to look at price extremums as one of the metrics of interest, volatility is still one of the core metrics for trading~\cite{bahmani2007exchange}.
To statistically analyse price changes and break down key events in the market Caginalp \& Caginalp~\cite{caginalp2018asset}, propose a method to find peaks in the volatility, leading to more prominent price levels.
But whilst this work is a recent approach, price levels were first used in the 80s and proven to be very successful~\cite{pruitt1988crisma, ferris1988predicting}
The price extrema represent the optimal point at which the price is being traded before a large fluctuation.
Since the implied relationship of supply and demand is something that will hold for any exchange, this is a great fit for various instruments.

In a different context, Miller et al~\cite{miller2019identification}, analyse Bitcoin data to find profitable trading bounds.
Bitcoin, unlike more traditional exchanges, is decentralised and traded 24h a day, making the data much more sparse and with less concentrated trading periods.
This makes the trends harder to analyse.
Their approach manipulates the data in such a way that it is smoothed, through the removal of splines, this seeks to manipulate curves to make its points more closely related.
By this technique, they can remove outliers and find clearer points of fluctuation as well as peaks.
The authors then construct a bounded trading strategy that proves to perform well against unbounded strategies.
Since Bitcoin has more decentralised access, and by the very nature of those investing in it, this also reduces barriers to entry, making automated trading much more common.
This indicates that techniques to identify bounds and points of interest in the market are also more favoured and widely used.

\subsubsection{Derivation of the market microstructure features}

A market microstructure is the study of financial markets and how they operate.
Its features represent the way that the market operates, how decisions are made about trades, the price discovery process and many more~\cite{kissell2013science}.
The process of market microstructure analysis is the identification of why and how the market prices will change, to trade profitably.
These may include, 1) the \textit{time between trades}, as it is usually  an indicator of trading intensity~\cite{bauwens2004comparison} 2) \textit{volatility}, which might represent evidence of good and bad trading scenarios, as high volatility may lead to an unsuitable market state~\cite{grammig2002modeling}, 3) \textit{volume}, which may directly correlate with trade duration, as it might represent informed trading rather than less high volume active trading~\cite{manganelli2005duration}, and 4) \textit{trade duration}, high trading activity is related to greater price impact of trades and faster price adjustment to trade-related events, whilst slower trades may indicate informed single entities~\cite{dufour2000time}.
In their work Munnix et al~\cite{munnix2012identifying}
proposed the characterisation of market structures based on correlation.
Through this, they were able to detect key states of market crises from raw market data.
Whist several other options are available they are often instrument-related and require expert domain knowledge.
In general, it is important to tailor and evaluate your features to cater to the specific scenario identified.

One such important scenario to consider when catering to prices is whether the price action is caused by aggressive buyers or sellers.
In an Order Book, a match implies a trade, which occurs whenever a bid matches an ask and conversely, however the trade is only ever initiated by one party.
To dictate who the \emph{aggressor} is in this scenario (if not annotated by the marketplace), the tick rule is used~\cite{aitken1996determinants}.
The rule labels a buy-initiated trade as 1, and a sell-initiated trade as -1.
The logic is the following an initial label $l$ is assigned an arbitrary value of 1 if a trade occurs and the price change is positive, then $l=1$ if the price change is negative, and $l=0$ and if there is no price change $l$ is inverted.
This can identify the aggressor with a high degree of accuracy~\cite{easley2016discerning}.

\subsection{Automated trading and strategies}

Rule-based methods have seen mixed successes in the context of financial time series~\cite{park2007we}.
In part, this is due to the wide usage of quite simple rules~\cite{cervello2015stock}.
What may instead be more desirable is to use patterns for trading~\cite{leigh2002analysis}.
These patterns can perform with high levels of profitability and can be automatically extracted from data, which is in itself, quite desirable.
Despite early criticisms~\cite{ratner1999tests}, not only have these patterns been shown to be particularly effective~\cite{cervello2015stock}, they might be particularly useful in the machine learning setting.

To evaluate the effectiveness of the trading strategy the Sharpe Ratio can be used.
This is a measure for assessing the performance of an investment or trading approach and can be computed as the following:
\begin{equation}
    S = \frac{R_p-R_f}{\sigma_p},
\end{equation}
where $R_p$ \& $R_f$ correspond to the portfolio and risk free returns, respectively, and $\sigma_p$ is a standard deviation of the portfolio return. 
While the equation gives a good intuition of the measure, in practice its annualised version is often computed. 
It assumes that daily returns follow the Wiener process distribution, hence to obtain annualised values, the daily Sharpe values are multiplied by $\sqrt{252}$ - the annual number of trading days. 
It should be noted that such an approach might overestimate the resulting Sharpe ratios as returns auto-correlations might be present, violating the original assumption~\cite{Lo2002TheRatios}.

Automating the trading through machine learning efforts is the identification of a market state in which a trade is profitable, and automatically performs the transaction at that stage.
Such a system is normally tailored for a specific instrument, analysing unique patterns to improve the characterisation.
One such effort focusing on Forex markets is, Dempster \& Leemans~\cite{dempster2006automated}.
In this work, a technique using reinforcement learning is proposed, to learn market behaviours.
Reinforcement learning is an area of machine learning concerned with how software agents ought to take actions in an environment to maximise the notion of cumulative reward.
This is achieved by assigning positive rewards to desired actions and negative rewards to undesired actions, leading to an optimisation towards actions that increment rewards. In financial markets, this naturally corresponds to profitable trades.
Using this approach, the authors can characterise when to trade, analyse associated risks, and automatically make decisions based on these factors.
    
\subsection{Backtesting}
\label{sec:backtest}
To test trading, strategy evaluation is performed to assess profitability. 
Whilst it is possible to do so on trading live market, it is more favourable to do so on the historical data first to get the results on a large timeline and avoid any financial risks.
The notion is that a strategy that would have worked poorly in the past will probably work poorly in the future, and conversely. 
But as you can see, a key part of backtesting is the weak assumption that past performance predicts future performance~\cite{harvey2015backtesting,bailey2014pseudomathematics,kahn1995does}.

Several approaches exist to perform backtesting and different things can be assessed.
Beyond trading strategies testing, backtesting can show how positions are opened and the likelihood of certain scenarios taking place within a trading period~\cite{harvey2015backtesting}.
The more common approach is to implement the backtesting within the trading software platform, as this has the advantage that the same code as live trading can be used.
Almost all platforms allow for simulations of historical data, although one operates on data pre-processed by the proprietary software platform, making the pipeline partially opaque.

For more flexibility, one can implement their backtesting system in languages such as C\# or Python~\cite{quantconnect}. 
This specific approach enables the same code pipeline that is training the classifier to also test the data, allowing for much smoother testing.
However, some limitation of this type of backtesting is that it doesn't include a lack of order book queues as well as the latency simulation, both of which are present in the live trading setting.
Consequently, slippages (the difference between where the order is submitted by the algorithm and the actual market entry/exit price) cannot be modelled accurately. 
Moreover, the modelling approaches and assumptions vary between the backtesting engines, hampering comparability. 
Nevertheless, backtesting is a commonly adopted practise among the financial market practitioners, and being used with care, gives a good estimation of the ATP quality. 

It is crucial to highlight that backtesting is only possible when all the components of the ATP are in place, making it barely suitable for assessing individual components separately.

\subsection{Machine learning algorithms in financial contexts}

There is a wide range of techniques for machine learning, ranging from very simple such as regression to techniques used for deep learning such as neural networks.
Consequently, it's important to choose an algorithm that is most suited to the problem one wishes to tackle. 
However, there is no uniform rule on how to make the most optimal choice and it is often empirically driven~\cite{kotthoff2012evaluation}.

In ATPs one of the possible roles of machine learning is to identify situations in which it is profitable to trade, depending on the strategy, for example, if using a flat strategy the intent is to identify when the market is flat.
In these circumstances, and due to the potentially high financial implications of false negatives understanding the prediction is key.
Understanding the prediction involves the process of being able to understand why the algorithm made this decision.
This is a non-trivial issue and something very difficult to do for a wide range of techniques, neural networks being the prime example (although current advances are being made~\cite{zhou2018interpreting,fan2020interpretability} .

\subsubsection{Algorithms}

Perhaps one of the simplest yet highly-effective techniques is known as Support Vector Machines (SVM)~\cite{pal2004assessment,janardhanan2015effectiveness}.
SVMs are used to identify the hyperplane that best separates a binary sampling.
If we imagine a set of points mapped onto a 2d plane, the SVM will find the best line that divides the two different classifications in half.
This technique can easily be expanded to work on higher-dimensional data, and since it is so simple, it becomes intuitive to see the reason behind a classification.
However, this technique whilst popular for usage in financial forecasting~\cite{tabak2009analysis,huang2005forecasting}, suffers from the drawback that it is very sensitive to parameter tuning, making it harder to use, and also does not work with categorical features directly, making it less suited to complex analysis~\cite{cawley2010over}.

Another popular approach is decision trees, as they have been around since the 1960s~\cite{magee1964decision}.
The reason tree-based approaches are hugely popular is that they are directly interpretable.
To improve the efficacy of this technique, several different trees are trained and used in unison to come up with the result.
A popular case of this is Random Forest~\cite{breiman2001random}.
Random Forest operates by constructing a multitude of decision trees at training time and outputting the classification of the individual trees.
However, this suffers from the fact that the different trees are not weighted and contribute equally, which might lead to inaccurate results~\cite{genuer2010variable}. Moreover, since the data is randomly sampled for individual trees, it is hard to expect the model to learn the most optimal data representation~\cite{genuer2010variable}.

One class of algorithms that have seen mass popularity for its increases in robustness, effectiveness and clarity is boosting algorithms~\cite{ridgeway1999state}.
Boosters create several classifications that can be combined to reduce overfitting and improve the prediction.
Initially, a weak classifier is trained, and then the subsequent classifier is trained with the additional weights of the initial classifier, putting further consideration on misclassified entries, consequently forcing the subsequent classifier to put more effort into classifying the results that are misclassified by the other classifier.  
After the set of classifiers is trained the results are aggregated with more weight being put on the more accurate learners~\cite{hastie2009elements}.

The first usage of boosting using a notion of weakness was introduced by AdaBoost~\cite{freund1997decision}, this work presented the concept of combining the output of the boosters into a weighted sum that represents the final output of the boosted classifier.
This allows for adaptive analysis as subsequent weak learners are tweaked in favour of those instances misclassified by previous classifiers.
Following on from this technique two other techniques were introduced XGBoost~\cite{chen2016xgboost} and LightGBM~\cite{ke2017lightgbm}, both libraries gained a lot of traction in the machine learning community for their efficacy, and are widely used.
In this category, the most recent algorithm is CatBoost.
CatBoost~\cite{prokhorenkova2018catboost} is highly efficient and less prone to bias than its predecessors, it is quickly becoming one of the most used approaches, in part due to its high flexibility.
CatBoost was specifically proposed to expand issues in the previous approaches which lead to target leakage, which sometimes led to overfitting.
This was achieved by using ordered boosting, a new technique allowing independent training and avoiding leakage~\cite{prokhorenkova2018catboost}.
This also allows for better performance on categorical features.

\subsubsection{Feature and prediction analysis}

Feature analysis is the evaluation of the input data to assess their effectiveness and contribution to the prediction.
This may also take the form of creating new features using domain knowledge to improve the data. 
The features in the data will directly influence the predictive models used and the results that can be achieved.
Intuitively, the better the chosen and prepared features, the better the results that can be achieved.
However, this may not always be the case for every scenario, as it may lead to overfitting due to a too large dimensionality of the data.
The process of evaluating and selecting the best features is referred to as feature selection~\cite{}.
One of the methods supporting feature engineering is feature importance evaluation.
The simplest way to achieve this is by feature ranking~\cite{guyon2003introduction}, in essence, a heuristic is chosen and each feature is assigned a score based on this heuristic, ordering the features in descending order. 
This approach however may be problem-specific and require domain knowledge.
Another common approach is the usage of correlations, to evaluate how the features relate to the output~\cite{yu2003feature}.
This approach intends to evaluate the dependency between the features and the result, which intuitively might lead to a feature that contributes more to the output~\cite{blessie2012sigmis}.
However, these approaches evaluate the feature as a single component, in relation to the output, independent of the other features.

Realistically one would want to understand their features as a whole and see how they contribute to a prediction as a group.
Compared to previously discussed approaches this starts from the result of the model and goes back to the feature to see which ones contributed to the prediction (referred to as post-hoc explanations)~\cite{adadi2018peeking}.
This has advantages over the pure feature analysis approaches as it can be applied to all the different predictors individually and gives insights into the workings of the predictor.
Recent advances in this approach, namely SHAP (SHapleyAdditive exPlanations)~\cite{lundberg2017unified}, can provide insight into a prediction scoring of each feature.
This innovative technique can allow the step-through assessment of features throughout the different predictions, providing guided insight which can also be averaged for an overall assessment.
This is very useful for debugging an algorithm, assessing the features and understanding the model decisions, making the approach particularly relevant for the current work.

\subsection{Statistical reproducibility}

To evaluate the results of our research, answer the research questions, add explainability and increase the reproducibility of our study, we make use of several statistical techniques.

\subsubsection{Effect sizes}

The first step that has to be done is quantifying the effectiveness of the approach in relation to control.
Statistically, this can be done using effect sizes~\cite{Durlak2009HowSizes}.
Effect size is a measure for calculating the strength of a statistical claim.
A larger effect size indicates that there is a larger difference between the treatment (method) and the control sample.
Reporting effect sizes is considered a good practice when presenting empirical research findings in many fields~\cite{stern2004good,hawkins2013statistical,wilkinson1999statistical}.
Two types of effect sizes exist: relative and absolute. Absolute ones provide a raw difference between the two groups and are usually used for quantifying the effect of a particular use case. Relatives are obtained by normalising the difference by the absolute value of the control group.
Depending on the setting, when computing effect sizes one might look at differences in variances explained, differences in mean, and associations in variables~\cite{kelley2012effect}.

Differences in variance explained assess the proportion to which a mathematical model accounts for the variation. 
One of the most common ways to do this is by making use of Pearson Correlation~\cite{ref1}. 
Pearson correlation is defined as the covariance of the two variables, divided by the product of their standard deviations.
This normalises the measurement of the covariance, such that the result always has a value between -1 and 1.
A further commonly used measure is known as r-squared, taking the Pearson correlation and squaring it.
By doing this, we can measure the proportion of variance shared by the two variables.
The second approach is instead to look at the differences in population means, using a standardisation factor.
Popular approaches include Cohen's \textit{d}~\cite{cohen1960coefficient}, which calculates the difference between two sample means with pooled standard deviation as the standardisation factor.
However, it was found that the standard deviation may be biased as the standardisation factor, meaning that when the two means are compared and standardised by division as follows $frac{u_1 - u_2}{SD}$, if the standard deviation $SD$ is used it may cause some bias and alternative standardisations may be preferred.
This is rectified in Hedge's \textit{g}~\cite{hedges1981distribution} method, which corrects the bias using a correction factor when computing the pooled standard deviation.
A further extension that can be added on top of this correction is to use $av$ or rather an average variance instead of variance, this is more powerful when applied to two correlated samples, and uses the average standard deviation instead, once again the corrected Cohen's $d_av$ is referred to as Hedge's $g_{av}$~\cite{cumming2013understanding,Lakens2013CalculatingANOVAs}.
The final type of effect size is categorical variable associations, which checks the inter-correlation of variables, and can evaluate the probability of variables being dependent on one another, examples of this are the chi-squared test~\cite{fraser2019association}, also effective on ordinal variables. 

\subsubsection{Inferential statistics}

Another core component of a statistical assessment is Inferential Statistics.
With inferential statistics, one aims to determine whether the findings are generalisable to the population.
It is also used to determine if there is a significant difference between the means of two groups, which may be related to certain features.
The most popular approaches fall under the general linear model~\cite{mcneil1996testing}.
The general concept is that we want to use a null hypothesis to test the probabilistic difference between our sample population and another population.
Popular approaches include t-test~\cite{student1908probable}, ANOVA~\cite{girden1992anova}, Wilcoxon~\cite{wilcoxon1992individual}, and many more, depending on the considered setting.
A t-test is a type of inferential statistic used to determine if there is a significant difference between the means of two groups, which may be related to certain features.
The basic functioning is that you take a sample from two groups, and establish the null hypothesis for which the sample means are equal, it then calculates the mean difference, standard deviations of the two groups and number of data values of two groups and attempts to reject the null hypothesis.
If the null hypothesis is rejected it means that the mean difference is statistically significant.
However the t-test relies on several assumptions: 1) that the data is continuous, 2) that the sample is randomly collected from the total populations, 3) that the data is normally distributed, and 4) that the data variance is homogeneous~\cite{usman2016consistency}.
This makes the t-test not suited to the analysis of small samples, where normality and other sample properties are hard to assess reliably.
An approach that doesn't face the same limitations is the Wilcoxon test~\cite{wilcoxon1992individual}.
The advantage of this approach is that instead of comparing means, it repeatedly compares the samples to check if the mean ranks differ, this means it will check the arithmetic average of the indexed position within a list.
This type of comparison is applicable for paired data only and done on individual paired subjects, increasing the power of the comparison.
However, a downside of this approach is that it is non-parametric.
A parametric test can better observe the full distribution and is consequently able to observe more differences and specific patterns, however as we saw with t-tests they rely on stronger assumptions and are sometimes impractical.

\subsubsection{Correction for multiple comparisons}

The more inferences are made, the more likely erroneous inferences are to occur.
Multiple comparisons arise when a statistical analysis involves multiple simultaneous statistical tests, each of which has the potential to produce the discovery, of the same dataset or dependent datasets. Hence, the overall probability of the discovery increases.
This increased chance should be corrected.
Some methods are more specific but there exist a class of general significance level $\alpha$ adjustments.
Examples of these are the Bonferroni Corrections~\cite{bonferroni1936teoria} and Šidák Corrections~\cite{vsidak1967rectangular}.
The general idea follows from the following: given that the p-value establishes that if the null hypothesis holds what is the likelihood of getting an effect at least as large in your sample.
Then if the p-value is small enough you can conclude that your sample is inconsistent with the null hypothesis and reject it for the population.
So the idea of the corrections is that to retain a prescribed significance level $\alpha$ in an analysis involving more than one comparison, the significance level for each comparison must be more stringent than the initial $\alpha$. 
In the case of Bonferroni corrections, if for some test performed out of the total $n$ we ensure that its p-value is less than $1.0-\alpha/n$, then we can conclude, as previously, that the associated null hypothesis is rejected.

\subsection{Methodologies for evaluation of ATP components}
\label{sec:backMethodATP}
    Below, we overview several works on trading pattern identification, design and trading from the methodological perspective. 
    
     In the work by Chen \& Chen~\cite{chen2016intelligent}, a pattern recognition model is proposed and evaluated for its performance. The performance is evaluated in two phases, model training and model testing. In the first phase, the authors perform optimisation of the pattern parameters to the current market conditions, in the second - perform pattern recognition and backtesting. 
     The same approach is taken in the work by Parracho et al~\cite{parracho2010trading}. 
    In the work by Cervello et al~\cite{cervello2015stock}, the authors thoroughly assess the proposed approach for trading pattern recognition based on ATP profitability for a range of take profits and stop losses. 
    
    An elegant approach to computational finance is proposed by Canelas et al~\cite{canelas2013sax}. It uses a symbolic approximation of the financial time series together with a genetic algorithm to find patterns and trading rules from the approximation. The results are reported in measures specific to this particular study, which do not allow direct comparison to other studies.  
    
    A statistical approach is taken in a paper published two decades ago~\cite{leigh2002analysis}. Namely, the authors introduce the null hypothesis in a form of the performance of a random-choice model and run a t-test on daily ATP profits to test the proposed method. This is a good first step in a statistically sound approach to financial markets. However, it falls short of justifying the choices within the methodology, reporting statistical results, and reporting measures that would allow the comparability. Moreover, the way the null hypothesis is formulated and evaluated limits the ability of a practitioner to test a particular ATP component, as discussed in Section~\ref{sec:backtest}.
    
    Concluding, there is no uniform dataset for reporting the results, and neither there is a uniform trading strategy. Moreover, all the considered studies assess multiple ATP components at once.
    The lack of uniformity and comparability highlights the demand for the current study.
    
\section{Material and methods}
\label{sec:methods}

\begin{figure}[!hbt]
        \centering
        \includegraphics[width=0.8\textwidth]{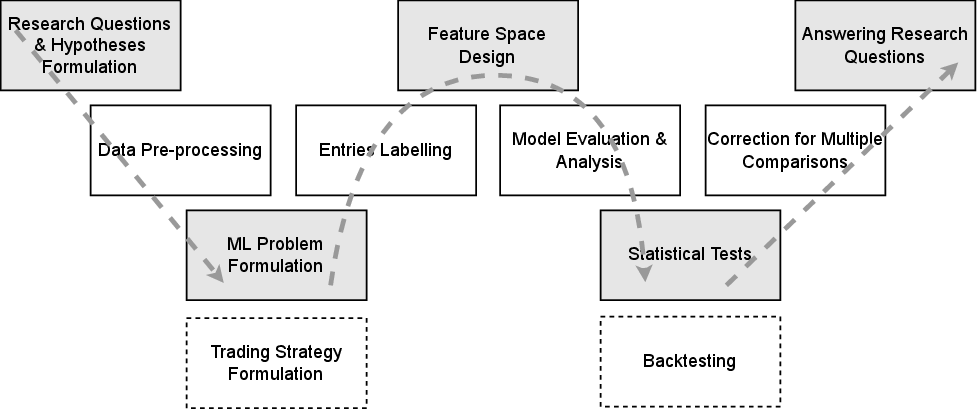}
        \caption{The methodology proposed for profitability, performance and statistical evaluation of the research questions in the field of financial markets. The bottom elements are optional and are not required to answer the RQs}
        \label{fig:methodology}
    \end{figure}

    In the section, we first describe the datasets and pre-processing procedures. 
    Then, the proposed methodology is applied (Figure~\ref{fig:methodology}), separately assessing the model performance and profitability, as well as statistically approaching the research questions and performing model analysis.
    
    The resulting experiment design comprises entries labelling and formulating the machine learning classification problem, designing features, evaluating model performance, evaluating its statistical significance, explaining the model and, finally, backtesting. 

\begin{figure}[!hbt]
        \centering
        \includegraphics[width=0.8\textwidth]{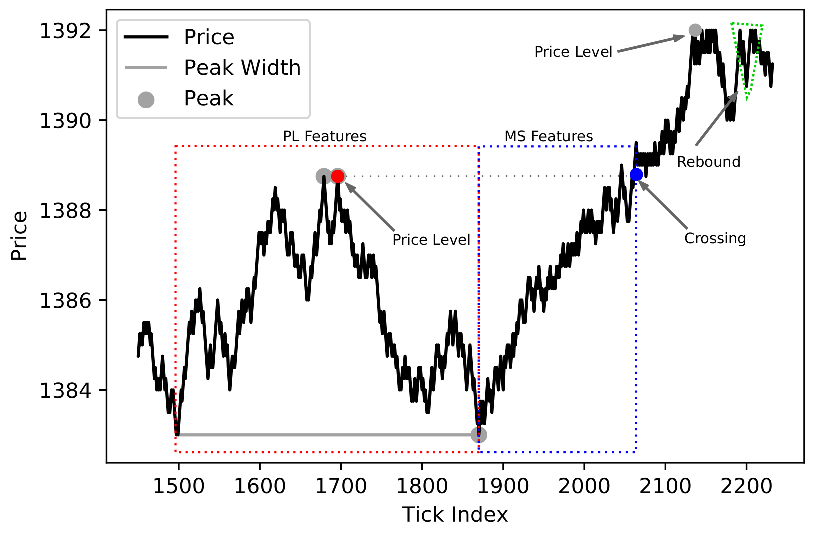}
        \caption{The figure illustrates what data we use for the 2-step feature extraction: price level (PL) and market shift (MS) components. It also demonstrates peaks, peak widths, as well as rebound and crossing labelling.}
        \label{fig:featTypes}
    \end{figure}

    We propose a price-action-based way of defining the price levels and performing their classification. Concretely: we identify local price extrema and predict whether the price will reverse (or 'rebound') or continue its movement (also called 'crossing').
    For the demonstration purposes, we set up a simplistic trading strategy, where we are trading a price reversal after a discovered local extremum is reached as shown in Fig.~\ref{fig:featTypes}. 
    We statistically assess our choices of the feature space and feature extraction method. 
    In the simulated trading, we limit our analysis to backtesting and do not perform any live trading. 
    In the \nameref{sec:discussion} section we address the limitations of such an approach. 
    Also, we share the reproducibility package for the study~\cite{zenodo2020paper}. It is implemented in Python, we also share the trained models. 

    \subsection{Data}
        In the study, we use S\&P500 E-mini CME futures contracts ES(H-Z)2017, ES(H-Z)2018 and ES(H-U)2019.
        Which correspond to ES futures contracts with expiration in March (H), June (M), September (U) and December (Z). We operate on ticks data which includes Time\&Sales records statistics, namely: bid and ask volumes and numbers of trades, as well as the largest trade volumes per tick.
        We consider a tick to incorporate all the market events between the two price changes by a minimum price step. 
        For the considered financial instrument the tick size is \$0.25. 
        
    \subsection{Data pre-processing}
        
        We sample the contract data to the active trading periods by considering only the nearest expiring contracts with the conventionally accepted rollover dates. The samples end on the second Thursday of the expiration month - on that day the active trading is usually transferred to the following contract. This decision ensures the highest liquidity, and, due to the double-auction nature of the financial markets, stable minimum bid-ask spreads~\cite{Iori2002AMarkets}.
        
        In the current study, we consider the two simplest scenarios of the price behaviour after it reaches the local extremum - reversal and extremum crossing. 
        When labelling the entries, we require up to 15 ticks of price movement as a reversal (or rebound) and only 3 ticks for the extremum crossing. The labelling approach allows us to study a range of configurations of the reversals and investigate how the configurations affect the performance of the models. At the same time, these ranges are well within the boundaries of the intraday price changes. 
        
        An essential part of the proposed automated trading system is the detection of the price extrema. 
        The detection is performed on a sliding window of the streamed ticks with a window size of 500. 
        We capture peaks with widths from 100 to 400. 
        The selected widths range serves three purposes: i) ensures that we do not consider high-frequency trading (HFT) scenarios which require more modelling assumptions and different backtesting engines; ii) allows us to stay in intraday trading time frames and have a large enough number of trades for analysis; iii) makes the price level feature values comparable across many of the entries.

        \subsection{Classification task}
            To incorporate machine learning into the automated trading system, we design a binary classification task, where the labels correspond to price reversals (positives) and crossings (negatives). Due to the labelling design, we are more flexible with taking profits and stopping losses when trading reversals (up to 15 ticks versus 3 ticks) - this explains their positive labelling. 
            
        \subsection{Feature design}
        
            To perform the extrema classification, we obtain two types of features: i) designed from the price level ticks (called price level (PL) features), and ii) obtained from the ticks right before the extremum is approached (called market shift (MS) features) as we illustrate in Fig.~\ref{fig:featTypes}.
            We believe (and statistically test it) that it is beneficial to perform the 2-step collection since the PL features contain properties of the extremum, and the MS features allow us to spot any market changes that happened between the time when the extremum was formed and the time we are trading it.
            
            Considering different extrema widths, varying dimensionality of the data does not allow to use it directly for classification - most of the algorithms take fixed-dimensional vectors as input. 
            We ensure the fixed dimensionality of a classifier input by aggregating per-tick features by price. 
            We perform the aggregation for the price range of 10 ticks below (or above in case of a minimum) the extremum. 
            This price range is flexible - 10 ticks are often not available within ticks associated with the price level (red dashed rectangle in Fig.~\ref{fig:featTypes}) in this case we fill the empty price features with zeros. 
            We assume that the further the price from the extremum the less information relevant for the classification it contains. 
            Considering the intraday volatility of ES, we expect that the information beyond 10 ticks from the extremum is unlikely to improve the predictions. 
            If one considers larger time frames (peak widths), this number might need to increase. 
    
            PL features are obtained from per-tick features by grouping by price with the \textit{sum}, \textit{max} or \textit{count} statistics. 
            For instance: if one is considering volumes, it is reasonable to \textit{sum} all the aggressive buyers and sellers before comparing them. Of course, one can also compute \textit{mean} or consider \textit{max} and \textit{min} volumes per tick.
            If following this line of reasoning, the feature space can be increased to very large dimensions. 
            We empirically choose a feature space described in Tables~\ref{tab:featSpacePL} and \ref{tab:featSpaceMS} for Price Level and Market Shift components, respectively. Defining the feature space we aim to make the feature selection step computationally feasible. Too large feature space might be also impractical from the optimisation point of view, especially if the features are correlated.

             \begin{table}[!htb]

            \caption{Price level feature space component was used in the study. These features are obtained when the price level is formed. When discussed, features are referred to by the codes in the square brackets at the end of descriptions.}
            
            \label{tab:featSpacePL}
            \centering 
            \scriptsize           
            \begin{tabular}{l|l}
            
                 \multicolumn{2}{l}{
                \begin{tabular}{|c|ll|}
                \hline
                \multirow{15}{*}{\centering\rotatebox{90}{\textbf{Price level (PL) features}}}& \textbf{Equation} & \textbf{Description} \\
                 \hline
                    &$\sum_{t}^{p=|PL-t|}(V_b+V_a)$ & \makecell[l]{ Bid and ask volumes summed across all the ticks for\\ $t\in[0,1,2]$ [PL0]} \\
                    &$\sum_{t}^{p=|PL-t|}V_b$ & Bid volumes summed across all the ticks for $t\in[0,1,2]$ [PL1] \\
                    &$\sum_{t}^{p=|PL-t|}V_a$ & Ask volumes summed across all the ticks for $t\in[0,1,2]$ [PL2] \\
                    &$\sum_{t}^{p=|PL-t|}T_b$ & \makecell[l]{ Number of bid trades summed across all the ticks for\\ $t\in[0,1,2]$ [PL3]} \\
                    &$\sum_{t}^{p=|PL-t|}T_a$ & \makecell[l]{Number of ask trades summed across all the ticks for \\ $t\in[0,1,2]$ [PL4]} \\
                    &$M(T{p=|PL-t|})_b$ & Maximum bid trade across all the ticks for $t\in[0,1,2]$ [PL5] \\
                    &$M(T_{p=|PL-t|})_a$ & Maximum ask trade across the ticks for $t\in[0,1,2]$ [PL6] \\
                    &$\sum_{t}^{p=|PL-t|}1$ & Number of ticks at price for $t\in[0,1,2]$ [PL7] \\
                    &$\frac{\sum_{t}^{p=|PL-t|}V_b}{\sum_{t}^{p=|PL-t|}V_a}$ & PL1 divided by PL2, for $t\in[0,1,2]$ [PL8] \\[8pt]
                    &$\frac{\sum_{t}^{p=|PL-t|}T_b}{\sum_{t}^{p=|PL-t|}T_a}$ & Feature PL3 divided by feature PL4 [PL9] \\[10pt]
                    &$\frac{\sum_{t}^{p=|PL-t|}M(T)_b}{\sum_{t}^{p=|PL-t|}M(T)_a}$ & Feature PL5 divided by feature PL6 [PL10] \\[8pt]
                    &$\frac{\sum_{t}^{p=|PL-t|}(V_b+V_a)}{\sum_{t}^{p=|PL-t|}1}$ & Total volume at price $|PL-t|$ divided by the number of ticks [PL11] \\[8pt]
                    &$\sum_{t=0}^{10}\sum_{t}^{p=|PL-t|}V_a$ & Total Ask Volume [PL12] \\[8pt]
                    &$\sum_{t=0}^{10}\sum_{t}^{p=|PL-t|}V_b$ & Total Bid Volume [PL13] \\[8pt]
                    &$\sum_{t=0}^{10}\sum_{t}^{p=|PL-t|}T_a$ & Total Ask Trades [PL14] \\[8pt]
                    &$\sum_{t=0}^{10}\sum_{t}^{p=|PL-t|}T_b$ & Total Bid Trades [PL15] \\[8pt]
                    &$\sum_{t=0}^{10}\sum_{t}^{p=|PL-t|}(V_a+V_b)$ & Total Volume [PL16] \\[8pt]
                    & - & Peak extremum - minimum or maximum [PL17] \\
                    & - & Peak width in ticks described in the Background section [PL18] \\
                    & - & Peak prominence - described in the Background section [PL19] \\
                    & - & Peak width height - described in the Background section [PL20] \\
                    & $\frac{\sum_{t\in[0,1,2]}^{p=|PL-t|}V_b}{\sum_{t\in[3..9]}^{p=|PL-t|}V_b}$ & \makecell[l]{ Bid volumes close to extremum divided by ones which are\\ further [PL21]} \\[6pt]
                    &$\frac{\sum_{t\in[0,1,2]}^{p=|PL-t|}V_a}{\sum_{t\in[3..9]}^{p=|PL-t|}V_a}$ & \makecell[l]{ Ask volumes close to extremum divided by ones which are \\further [PL22] }\\[6pt]
                    & $\frac{\sum_{t\in[0,1,2]}^{p=|PL-t|}V_b}{\sum_{t\in[0,1,2]}^{p=|PL-t|}V_a}$ &  \makecell[l]{Sum bid volume close to the price extremum divided by\\ the close ask volume [PL23]} \\[8pt]
                    &$\frac{\sum_{t\in[3..9]}^{p=|PL-t|}V_b}{\sum_{t\in[3..9]}^{p=|PL-t|}V_a}$ & \makecell[l]{Sum bid volume far from the price extremum divided by\\ the far ask volume [PL24]}\\
            
                \hline
                \hline
                 \multirow{ 1}{*}{\centering\rotatebox{90}{\textbf{Key}}}
                & \multicolumn{2}{l|}{
                        \begin{tabular}{lllll}
                          OB - order book & T - trades & t - ticks & N - total ticks &
                          p - price \\ w - tick window & PL - extremum price & V - volume &
                            b - bid & a - ask \\ \multicolumn{2}{l}{$P_{N}$ - price level neighbours until distance }  &   \multicolumn{3}{l}{M(X) - Max value in set X} 
                        \end{tabular}
                        }\\
                    \hline
                    \end{tabular}
                }\\
                
                \end{tabular}
            \end{table}
            \begin{table}[!htb]
            \footnotesize
                \centering
                \scriptsize
                \caption{Market shift feature space component used in the study. These features are obtained right before the already formed price level is approached. When discussed, features are referred to by the codes in the square brackets at the end of descriptions.}
                \label{tab:featSpaceMS}
                \begin{tabular}{l|l}
                     \multicolumn{2}{l}{
                    \begin{tabular}{|r|ll|}
                    \hline
                    \multirow{15}{*}{\centering\rotatebox{90}{\textbf{Market Shift (MS) features}}}& \textbf{Equation} & \textbf{Description} \\
                     \hline
                        &$\frac{ \sum_{t,b}^{w=237}(V_b)}{\sum_{t,a}^{w=237}(V_a)}$ & Fraction of bid over ask volume for last 237 ticks [MS0] \\[10pt]
                        &$\frac{ \sum_{t,b}^{w=237}(T_b)}{\sum_{t,a}^{w=237}(T_a)}$ &  Fraction of bid over ask trades for last 237 ticks [MS1] \\[10pt]
                        &$\frac{ \sum_{t}^{w=237}V_b}{\sum_{t}^{w=237}V_a} - \frac{\sum_{t}^{w=21}V_b}{\sum_{t}^{w=21}V_a}$ & \makecell[l]{Fraction of bid/ask volumes for long  minus\\ short periods [MS2]} \\[10pt]
                        &$\frac{ \sum_{t}^{w=237}T_b}{\sum_{t}^{w=237}T_a} - \frac{\sum_{t}^{w=21}T_b}{\sum_{t}^{w=21}T_a}$ & Fraction of bid/ask trades for long minus short periods [MS3] \\[10pt]
                        &$\frac{ \sum_{t}^{w=237} M(T)_b}{\sum_{t}^{w=237} M(T)_a} - \frac{\sum_{t}^{w=21} M(T)_b}{\sum_{t}^{w=21} M(T)_a}$ & \makecell[l]{Max bid trade divided by ask  for long periods minus \\ short periods [MS4]} \\[10pt]
                        &RSI($z\in S$) & Technical indicator RSI with the stated periods $z$ [$MS5_X$] \\[6pt]
                        &MACD($lp\in S$; $sp=lp/2$) & \makecell[l]{Technical indicator MACD with the stated long \& \\ short periods  $lp,sp$ [$MS6_X$]} \\[10pt]
                        \hline
                        \hline
                         \multirow{ 1}{*}{\centering\rotatebox{90}{\textbf{Key}}}
                        & \multicolumn{2}{c|}{
                                \begin{tabular}{llll}
                                   T - trades &   N - total ticks & w - tick window  & V - volume \\
                                    b - bid & a - ask & \multicolumn{2}{l}{$P_{N}$ - price level neighbours until distance } \\   \multicolumn{2}{l}{M(X) - Max value in set X} & \multicolumn{2}{l}{S - ranges [20,40,80,120,160,200]}\\
                                    t - ticks & z - time period & &
                                \end{tabular}
                                }\\
                            \hline
                            \end{tabular}
                        }\\
                
                \end{tabular}
            \end{table}

            To track the market changes, for the MS feature component we use 237 and 21 ticks and compare statistics obtained from these two periods. Non-round numbers help avoid interference with the majority of manual market participants who use round numbers~\cite{de2018advances}. 
            We also choose the values to be comparable to our expected trading time frames. No optimisation was made on them.
            We obtain the MS features being 2 ticks away from the price level to ensure that our modelling does not lead to any time-related bias where one cannot physically send the order fast enough to be executed on time.
            
        \subsection{Model evaluation}

            After the features are designed and extracted, the classification can be performed. 
            As a classifier, we choose the CatBoost estimator. 
            We feel that CatBoost is a good fit for the task since it is resistant to over-fitting, stable in terms of parameter tuning, efficient and one of the best-performing boosting algorithms~\cite{prokhorenkova2018catboost}. 
            Finally, being based on decision trees, it is capable of correctly processing zero-padded feature values when no data at price is available. 
            Other types of estimators might be comparable in one of the aspects and require much more focus in the other ones. 
            For instance, operating on a relatively small number of tabular data entries with a temporal component, we withhold from applying deep learning models due to evidence suggesting that boosting trees works better off-the-shelf in the considered setting~\cite{borisov2021deep}.
            
            In this study we use precision as the main scoring function (\textit{S}):
    
            \begin{equation}
                S=\frac{TP}{TP+FP},
            \label{eq:precision}
            \end{equation}
            where \textit{TP} is is the number of true positives and \textit{FP} the number of false positives. All the statistical tests are run on the precision scores of the samples. 
            This was chosen as the main metric since by design every \textit{FP} leads to losses, and a false negative (\textit{FN}) means only a lost trading opportunity. 
            To give a more comprehensive view of the model performance, we build confusion matrices, and compute F1 scores, PR-AUC (precision-recall area under the curve) and ROC-AUC (receiver-operating characteristic area under the curve) metrics. 
            We report model performances for the 2-step feature extraction approach as well as for each of the feature extraction steps separately.

            To avoid large bias in the base classifier probability, we introduce balanced class weights into the model. The weights are inversely proportional to the number of entries per class.
            The contracts for training and testing periods are selected sequentially - training is done on the active trading phase of contract $N$, testing - on $N+1$, for $N\in[0,B-1]$, where $B$ is the number of contracts considered in the study.
            
            We apply a commonly accepted ML community procedure for input feature selection and model parameter tuning~\cite{Kuhn2019FeatureModels}.
            Firstly, we perform the feature selection step using a Recursive Feature Elimination with the cross-validation (RFECV) method. 
            The method is based on the gradual removal of features from the model input starting from the least important ones (based on the model's feature importance) and measuring the performance on a cross-validation dataset. In the current study on each RFECV step, we remove 10\% of the least important features.
            Cross-validation allows robust assessment of how the model performance generalises into unseen data. Since we operate on the time series, we use time series splits for cross-validation to avoid the look-ahead bias. For the feature selection, the model parameters are left default, the only configuration we adjust is class labels balancing as our data is imbalanced. 
            Secondly, we optimise the parameters of the model in a grid-search fashion.
            Even though CatBoost has a very wide range of parameters that can be optimised, we choose the parameters common for boosting tree models for the sake of feasibility of the optimisation and leaving the possibility of comparing the optimisation behaviour to the other boosting algorithms. The following parameters are optimised: 1) \textit{Number of iterations}, 2) \textit{Maximum depth of trees}, 3) \textit{has\_time parameter set to True or False}, and 4) \textit{L2 regularisation}. 
            For the parameter optimisation, we use a cross-validation dataset as well. 
            We perform training and cross-validation within a single contract and the backtesting of the strategy on the subsequent one to ensure the relevance of the optimised model.

        \subsection{Statistical evaluation}
        Here we formalise the research questions by proposing null and alternative hypotheses, suggesting statistical tests for validating them, as well as highlighting the importance of the effect sizes. 
        
        The effect sizes are widely used in empirical studies in social, medical and psychological sciences~\cite{Lakens2013CalculatingANOVAs}. They are a practical tool allowing to quantify the effect of the treatment (or a method) in comparison to a control sample. Moreover, they allow the generalisation of the findings to the whole population (unseen data in our case). Finally, effect sizes can be compared across studies~\cite{Lakens2013CalculatingANOVAs}. We believe that introduction of the effect sizes into the financial markets domain contributes to the research reproducibility and comparability. 
            
        In the current study, we report Hedge's \textit{g}$_{av}$ - an unbiased measure designed for paired data. In the supplementary data, the effect sizes are provided in a form of forest plots with .95 confidence intervals (CIs), representing the range where the effect size for the population might be found with the .95 probability. We correct the confidence intervals for multiple comparisons by applying Bonferroni corrections. 
        
        When testing the hypotheses, the samples consist of the test precisions on the considered contracts, leading to equal sample sizes in both groups, and entries are paired as the same underlying data is used. Comparing a small number of paired entries and being unsure about the normality of their distribution, we take a conservative approach and for hypothesis testing use the single-tailed Wilcoxon signed-rank test. This test is a non-parametric paired difference test, which is used as an alternative to a t-test when the data does not fulfil the assumptions required for the parametric statistics. When reporting the test outcomes, we support them with the statistics of the compared groups. Namely, we communicate standard deviations, means and medians. 
            
        We set the significance level of the study to $\alpha=.05$. Also, we account for multiple comparisons by applying Bonferroni corrections inside of each experiment family~\cite{ce1936teoria}. We consider research questions as separate experiment families. 
            
        \subsection{Model analysis}
            
           We perform the model analysis in an exploratory fashion - no research questions and hypotheses are stated in advance. Hence, the outcomes of the analysis might require additional formal statistical assessment.
           We use SHAP local explanations to understand how models end up with the particular outputs. Through the decision plot visualisations, we aim to find common decision paths across entries as well as informally compare models with small and large numbers of features.   
           The reproducibility package contains the code snippets as well as the trained models, which allow repeating the experiments for all the models and contracts used in the study. 

        \subsection{Backtesting}
\begin{figure}[!htb]
                    \centering
                    \includegraphics[width=0.5\textwidth]{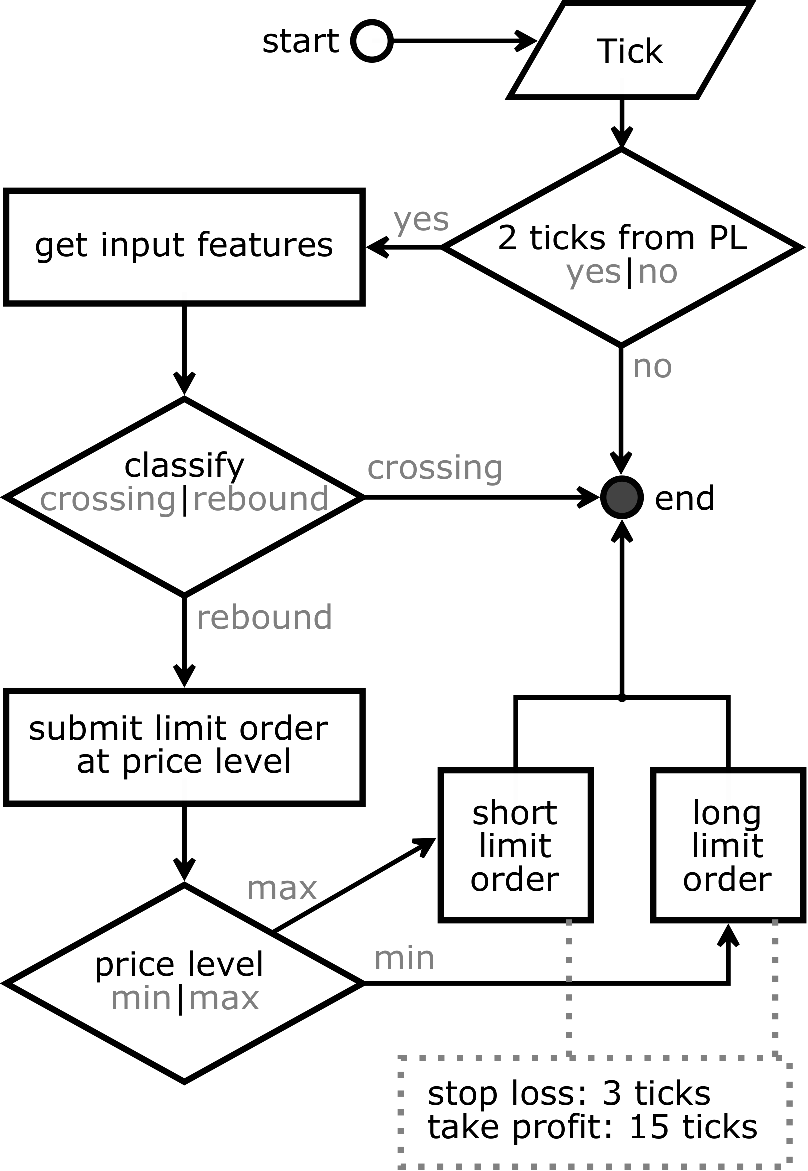}
                    \caption{The block diagram illustrates the steps of the trading strategy.}
                    \label{fig:strata}
            \end{figure}

            The trading strategy is defined based on our definition of the crossed and rebounded price levels, and schematically illustrated in Fig.~\ref{fig:strata}. 
            It is a flat market strategy, where we expect a price reversal from the price level.
            Backtrader Python package~\footnote{Available at: https://www.backtrader.com/} is used for backtesting the strategy.  
            Backtrader does not allow taking bid-ask spreads into account, that is why we are minimising its effects by excluding HFT trading opportunities (by limiting peak widths) and limiting ourselves to the actively traded contracts only. Since ES is a very liquid trading instrument, its bid-ask spreads are usually 1 tick, which however does not always hold during extraordinary market events, scheduled news, or session starts and ends. We additionally address the impact of spreads as well as order queues in the \nameref{sec:discussion} section. 

            In our backtests, we evaluate the performance of the models with different rebound configurations and fixed take-profit parameters, and varying take-profits with a fixed rebound configuration to better understand the impact of both variables on the simulated trading performance.

    \subsection{Approaching research questions}
        In the current subsection, we continue formalising the research questions by proposing the research hypotheses. The hypotheses are aimed to support the findings of the paper, making them easier to communicate. For both research questions, we run the statistical tests on the precision metric, as justified above.
        
        In the current study, we encode the hypotheses in the following way: H$_{0X}$ and H$_{1X}$ correspond to null and alternative hypotheses, respectively, for research question X.
        
        \paragraph{CatBoost versus no-information model (RQ1)}
            In the first research question, we investigate whether it is feasible to improve the baseline performance for the extrema datasets using the chosen feature space and CatBoost classifier. 
            We consider the baseline performance to be the precision of an always positive class output estimator.
            The statistical test is performed on the following hypothesis:
            
            \textbf{H$_{11}$:} CatBoost estimator allows classification of the extracted extrema with a precision better than the no-information approach.
            
            \textbf{H$_{01}$:} CatBoost estimator allows classification of the extracted extrema with a worse or equal precision in comparison to the no-information approach.
            
        \paragraph{2-step versus single-step feature extraction (RQ2)}
            The second research question assesses if the proposed 2-step feature extraction gives any statistically significant positive impact on the extremum classification performance. 
            The statistical test addresses the following hypothesis:
            
            \textbf{H$_{12}$:} 2-step feature extraction leads to an improved classification precision in comparison to using features extracted from any of the steps on their own.
            
            \textbf{H$_{02}$:} 2-step feature extraction gives equal or worse classification precision than features extracted from any of the steps on their own. 
            \newline
            In the current setting, we are comparing the target sample (the 2-step approach) to the two control samples (MS and PL components). We are not aiming to formally relate the MS and PL groups, hence only comparisons to the 2-step approach are necessary. To reject the null hypothesis, the test outcomes for both MS and PL components should be significant.

\section{Results}
\label{sec:results}
    In the current section, we communicate the results of the study. Namely: the original dataset and pre-processed data statistics, model performance for all the considered configurations, statistical evaluation of the overall approach and the 2-step feature extraction, and, finally, simulated trading and model analysis. 
    An evaluation of these results is presented in Sec.~\ref{sec:discussion}. Further analysis is provided in  \nameref{sec:supplementary} for completeness.
    
    \subsection{Raw and pre-processed data}
        The considered data sample and processed datasets are provided in Tab.~\ref{tab:processedEntries}. We provide the numbers of ticks per contract in the first columns. The contracts are sorted by the expiration date from top to bottom ascendingly. The number of ticks changes non-monotonically - while the overall trend is rising, the maximum number of ticks is observed for the ESZ2018 contract. And the largest change is observed between ESZ2017 \& ESH2018. 

        We perform the whole study on 3 different rebound configurations: 7, 11 and 15 ticks price movement required for the positive labelling. We communicate the numbers of positively labelled and total numbers of entries used in the classification tasks in Table~\ref{tab:processedEntries}. As one can see, the numbers of the extracted extrema do not strictly follow the linear relation with the numbers of ticks per contract. Considering the numbers of positively labelled entries, the numbers decrease for the larger rebound sizes. 
    
        \begin{table}[!htb]
            \centering
            \scriptsize
            \caption{The table communicates the numbers of reconstructed ticks per contract, numbers of positive labels, as well as total numbers of entries per contract. 'Reb.' corresponds to the rebound - the required number of ticks for the positive labelling. Rebound columns show numbers of positively labelled entries.}
                \begin{tabular}{l|r|rrr|r}
               Contract & No. Ticks &Reb. 7 & Reb. 11 & Reb. 15 &  Total \\
               \hline
               ESH2017 &  1271810 &   896 &   804 &   703 &   4911 \\
               ESM2017 &  1407792 &   951 &   881 &   756 &   4181 \\
               ESU2017 &  1243120 &   858 &   812 &   693 &   3446 \\
               ESZ2017 &  1137427 &   689 &   640 &   518 &  12317 \\
               ESH2018 &  2946336 &  2014 &  1983 &  1953 &  11537 \\
               ESM2018 &  2919757 &  1965 &  1936 &  1868 &   6534 \\
               ESU2018 &  1825417 &  1271 &  1226 &  1095 &  16331 \\
               ESZ2018 &  3633969 &  2677 &  2620 &  2565 &  11718 \\
               ESH2019 &  3066530 &  1990 &  1949 &  1883 &   9743 \\
               ESM2019 &  2591000 &  1711 &  1692 &  1630 &  10389 \\
               ESU2019 &  2537197 &  1761 &  1735 &  1704 &   6191 \\
                \end{tabular}
                \label{tab:processedEntries}
        \end{table}

    \subsection{Price Levels}
        
        \subsubsection{Automatic extraction}
            The first step of the pipeline was detecting peaks, which is done automatically, the same way as shown in Fig.~\ref{fig:featTypes}. We mark peaks with grey circles and the associated peak widths are depicted with solid grey lines. In our setting some of the peaks are not automatically discovered as they were not satisfying the conditions of the algorithm by having insufficient widths or being not prominent enough (see Section~\ref{sec:background:PriceExtrema} for the definitions of both).

        \subsubsection{Classification of the extrema}
        
\begin{figure}[!htb]
        \centering
        \includegraphics[width=0.8\textwidth]{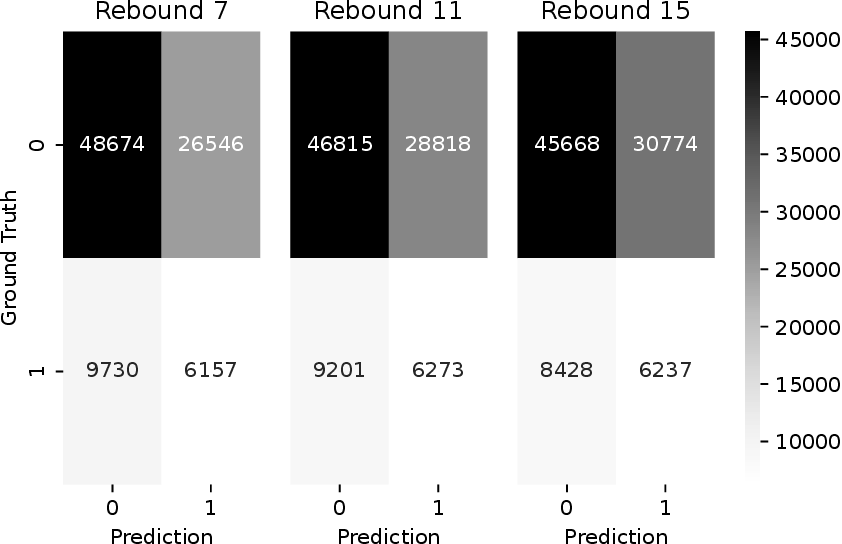}
        \caption{Confusion matrices for all the rebound configurations, built for the 2-step feature extraction method.}
        \label{fig:confMatx}
    \end{figure}

            For all the considered models we performed feature selection and parameter tuning. We made the optimisation results available as a part of the reproducibility package.  
            The model precision obtained on a per-contract basis for the 2-step feature extraction, PL and MS feature spaces, and no-information model is reported in Table~\ref{tab:RQ12MSPprecs}. The relative changes in the precision across contracts are preserved across the labelling configurations. We also provide the confusion matrices for the 2-step feature extraction as aggregate performance across all the contracts in Figure~\ref{fig:confMatx}. 
            
            We report the rest of the metrics for the 2-step feature extraction method, PL and MS feature in the supplementary data, in Tables~\ref{tab:AllMetricsPSuppl},~\ref{tab:AllMetricsPSuppl},~\ref{tab:AllMetricsMSSuppl}, respectively.

            \begin{table}[!htb]
                \centering
                \scriptsize
                \caption{Precisions of CatBoost classifier with the 2-step feature extraction ('2-step') and always-positive output classifier ('Null'). As well as Precisions of the Market Shift (MS) and Price Level (PL) feature spaces. For 7,11 and 15 ticks rebounds.}
                    \begin{tabular}{l|rrrr|rrrr|rrrr}
                Contract &  \multicolumn{4}{|c|}{Rebound 7} &  \multicolumn{4}{c|}{Rebound 11} & \multicolumn{4}{c}{Rebound 15} \\
                \hline                                                                                                                                       
                {} &          2-step &  Null             &   MS &  PL       &  2-step &             Null   &    MS &  PL        &  2-step &  Null \           &   MS &  PL       \\ \hline                                     
                ESH2017 &       .20 &            .19   &  .20 &      .19&       .20 &            .18 &  .19 &       .17 &       .15 &            .15& .15 &       .15\\
                ESM2017 &       .22 &            .21   &  .24 &      .21&       .23 &            .19 &  .20 &       .19 &       .19 &            .17& .16 &       .18\\
                ESU2017 &       .25 &            .20   &  .25 &      .23&       .15 &            .19 &  .23 &       .17 &       .18 &            .15& .15 &       .19\\
                ESZ2017 &       .17 &            .16   &  .18 &      .16&       .17 &            .16 &  .16 &       .16 &       .19 &            .16& .16 &       .15\\
                ESH2017 &       .19 &            .17   &  .19 &      .17&       .18 &            .17 &  .18 &       .18 &       .17 &            .16& .17 &       .16\\
                ESM2018 &       .21 &            .19   &  .22 &      .19&       .21 &            .19 &  .21 &       .20 &       .18 &            .17& .18 &       .17\\
                ESU2018 &       .17 &            .16   &  .17 &      .15&       .16 &            .16 &  .17 &       .16 &       .16 &            .16& .16 &       .15\\
                ESZ2018 &       .18 &            .17   &  .18 &      .17&       .17 &            .17 &  .17 &       .18 &       .16 &            .16& .17 &       .16\\
                ESH2019 &       .18 &            .18   &  .19 &      .18&       .18 &            .17 &  .18 &       .17 &       .18 &            .17& .17 &       .17\\
                ESM2019 &       .18 &            .17   &  .18 &      .18&       .17 &            .17 &  .18 &       .17 &       .17 &            .16& .18 &       .16\\
                    \end{tabular}
                    \label{tab:RQ12MSPprecs}
        \end{table}
 
        \subsubsection{Price Levels, CatBoost versus No-information estimator (RQ1)}
            Below we present the effect sizes with .95 confidence intervals (CIs) (Table~\ref{tab:RQ12testStats}) associated with the research question. Concretely, we use precision as the measurement variable for comparing the no-information model and the CatBoost classifier. No configurations were showing significant effect sizes. The largest effect size was observed for the 15 tick rebound labelling. Large CIs were observed partially due to the small sample size.

        We tested the null hypothesis for rejection for the 3 considered configurations. The original data used in the tests are provided in Table~\ref{tab:RQ12MSPprecs} - '2-step' and 'Null' columns. We reported test outcomes in a form of test statistics and p-values in Table~\ref{tab:RQ12testStats}. Additionally, in the same table, the sample statistics are included. We illustrate the performance of the compared groups in the supplementary materials, in Figure~\ref{fig:RQ1PrecSuppl}. There was no skew in any of the labelling configurations - medians and means do not differ within the groups. We saw around 2 times larger standard deviations for the CatBoost model in comparison to the no-information model. The potential reasons and implications are discussed in sections~\ref{sec:Discussion:RQ1} and~\ref{sec:Discussion:Limitations}, respectively. 
        There are 3 tests run in this experiment family, hence after applying Bonferroni corrections for multiple comparisons, the corrected significance level was $\alpha=.05/3=.0167$.

        \subsubsection{Price Levels, 2-step feature extraction versus its components (RQ2)}
        
            We display the effect sizes related to the second research question in Table~\ref{tab:RQ12MSPprecs} - '2-step', 'PL', 'MS' columns. We report them separately for PL and MS components versus the 2-step. There were no significant effects observed for any of the labelling configurations. Moreover, for the considered sample MS effect sizes were negative for rebounds 7 and 11 (Tab.~\ref{tab:RQ12testStats}). The negative effect size in the considered setting meant that the MS compound performs better than the 2-step approach. This effect was insignificant and hence does not generalise to the population.

            We performed statistical tests to check if the null hypothesis H$_{02}$ could be rejected. The original data used in the tests are provided in Table~\ref{tab:RQ12MSPprecs}, for the target (2-step) and control groups (MS and PL), respectively. We communicate the test outcomes in Table~\ref{tab:RQ12testStats}. For the sake of reproducibility, in the same table, we reported the compared groups' standard deviations, means and medians. We interpreted the results of the tests in section~\ref{sec:Discussion:RQ2}. Finally, to support the reader, we plot the performance of the considered groups in the supplementary data, Figure~\ref{fig:RQ2PrecSuppl}. There are 6 tests run in this experiment family, hence after applying Bonferroni corrections, the corrected significance level was $\alpha=.05/6=.0083$.

\begin{table}[!htb]
    \caption{Statistics supporting the outcomes of the Wilcoxon test which assessed whether the CatBoost estimator with the 2-step feature extraction ('CB' column) lead to a better classification performance than the always-positive output classifier ('Null' column) and whether the CatBoost estimator with the 2-step feature extraction ('2-step' column) lead to a better classification performance than each of the single-step feature extraction ('PL' and 'MS' columns) . The result is reported for the rebound labelling configurations of 7, 11 and 15 ticks.}
    \centering
    \scriptsize
    \begin{tabular}{c|c|c||c|c|c|c}
    {Statistics} & \multicolumn{6}{c}{Test Groups}\\
    \hline

   & \multicolumn{6}{c}{\textbf{Rebound 7}} \\

   & \multicolumn{2}{c}{\textbf{RQ1}} &  \multicolumn{4}{c}{\textbf{RQ2}}  \\   \hline
  \multicolumn{3}{c}{\textbf{}} & \multicolumn{2}{c}{\textbf{PL}} &  \multicolumn{2}{c}{\textbf{MS}}  \\   \hline
       One-tailed Wilcoxon test p-value & \multicolumn{2}{c||}{$<.001$} &  \multicolumn{2}{c}{.0049} & \multicolumn{2}{c}{.96}\\
       Test Statistics & \multicolumn{2}{c||}{55.0}  & \multicolumn{2}{c}{52.0} & \multicolumn{2}{c}{11.0}\\
       Effect Size (Hedges \textit{g$_{av}$}) & \multicolumn{2}{c||}{$0.64\pm1.02$}  & \multicolumn{2}{c}{$0.57\pm1.15$} & \multicolumn{2}{c}{$-0.1\pm1.07$}\\
       \hline
             & CB & Null &  2-step & PL & 2-step & MS \\
       Mean (Precision) & .19 & .18&  .19 & .18 & .19 & .20 \\
       Median (Precision) & .18 & .17& .18 & .18 & .18 & .19\\
       Standard Deviation (Precision) & .025 & .0151  & .025 & .018 & .025 & .028\\

    \end{tabular}

    \begin{tabular}{c|c|c||c|c|c|c}
    \hline
    & \multicolumn{6}{c}{\textbf{Rebound 11}} \\
    
   & \multicolumn{2}{c}{\textbf{RQ1}} &  \multicolumn{4}{c}{\textbf{RQ2}}  \\   \hline
   \multicolumn{3}{c}{\textbf{}} & \multicolumn{2}{c}{\textbf{PL}} &  \multicolumn{2}{c}{\textbf{MS}}  \\   \hline
       One-tailed Wilcoxon test p-value & \multicolumn{2}{c||}{.053} &  \multicolumn{2}{c}{.080} & \multicolumn{2}{c}{.58}\\
       
       Test Statistics & \multicolumn{2}{c||}{44.0}  & \multicolumn{2}{c}{42.0} & \multicolumn{2}{c}{26.0}\\
        Effect Size (Hedges \textit{g$_{av}$}) & \multicolumn{2}{c||}{$0.38\pm0.96$}  & \multicolumn{2}{c}{ $0.6\pm1.16$} & \multicolumn{2}{c}{$-0.02\pm1.06$}\\
       \hline
             & CB & Null & 2-step & PL & 2-step & MS \\
       Mean (Precision) & .18 & .17& .18 & .18 & .18 & .19\\
       Median (Precision) & .18 & .17 & .18 & .18 & .18 & .18\\
       Standard Deviation (Precision) & .022 & .0112  & .022 & .012 & .022 & .017\\

    \end{tabular}

    \begin{tabular}{c|c|c||c|c|c|c}
    \hline
    & \multicolumn{6}{c}{\textbf{Rebound 15}} \\
    
   & \multicolumn{2}{c}{\textbf{RQ1}} &  \multicolumn{4}{c}{\textbf{RQ2}}  \\   \hline
  \multicolumn{3}{c}{\textbf{}} & \multicolumn{2}{c}{\textbf{PL}} &  \multicolumn{2}{c}{\textbf{MS}}  \\   \hline
       One-tailed Wilcoxon test p-value & \multicolumn{2}{c||}{.0049} & \multicolumn{2}{c}{.052} & \multicolumn{2}{c}{.35}\\
       Test Statistics & \multicolumn{2}{c||}{52.0} & \multicolumn{2}{c}{44.0} & \multicolumn{2}{c}{32.0}\\
       Effect Size (Hedges \textit{g$_{av}$}) & \multicolumn{2}{c||}{$1.05\pm1.16$}  & \multicolumn{2}{c}{$0.8\pm1.22$} & \multicolumn{2}{c}{$0.2\pm1.07$}\\
       \hline
             & CB & Null & 2-step & PL & 2-step & MS \\
       Mean (Precision) & .17 & .16&  .17 & .16 & .17 & .17\\
       Median (Precision) & .17 & .16&  .17 & .16 & .17 & .17\\
       Standard Deviation (Precision) & .012 & .0055 & .012 & .006 & .012 & .010\\
    \end{tabular}
    
    \label{tab:RQ12testStats}
\end{table}

        \subsection{Model analysis}
            Here we show the exploratory analysis of the trained models. We choose two models trained on the same contract but with different labelling configurations. Namely, we report the analysis of the models trained on the ESH2019 contract, with rebounds 7 and 11. The choice is motivated by very different numbers of features after the feature selection step.

            \begin{figure}[!htb]
                \centering

                    \includegraphics[width=0.7\textwidth]{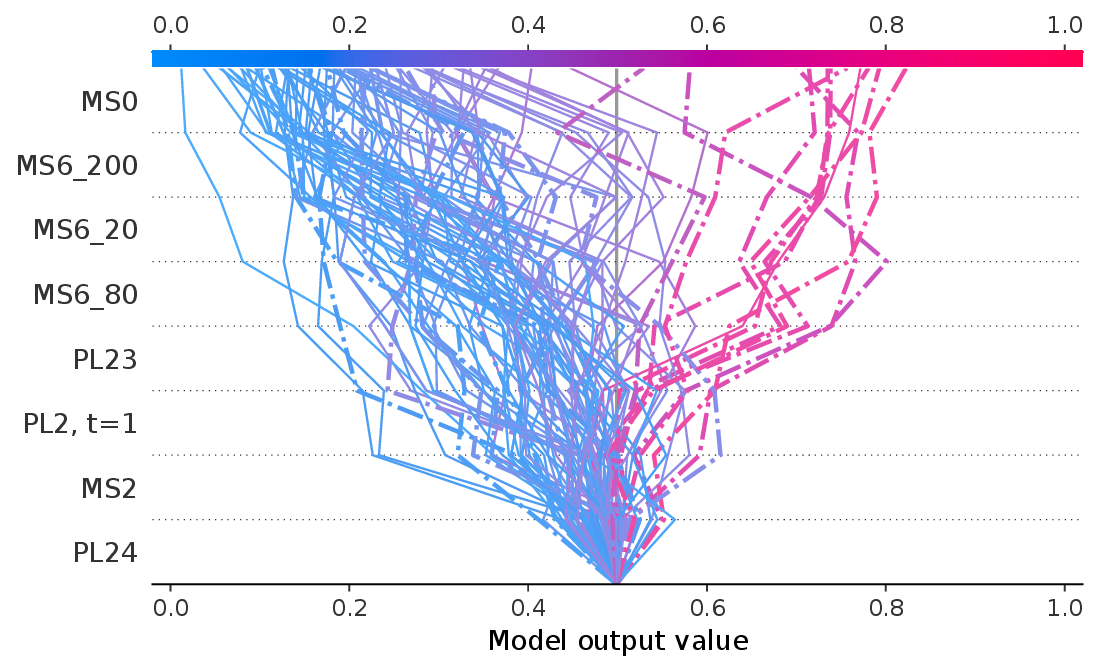}
                    \caption{Feat. contributions to the output on a per-entry basis for the CatBoost model, trained on ESH2019, rebound 7 configuration. The X-axis shows the strength of the contribution either towards a positive class (when the change is \textgreater 0) or towards the negative one. blue corresponds to the negative class (crossing) and red - to the positive (rebound). Misclassified entries are depicted with dashed lines.}
                    \label{fig:decPlotESH2019reb7}
            \end{figure}

    The core analysis is done on the decision plots, provided in Fig.~\ref{fig:decPlotESH2019reb7}.
    Since no pattern was observed when plotting the whole sample, we illustrate a random subsample of 100 entries from the top 1000 entries sorted by the per-feature contribution. 
       There are 29 features in the model trained on the rebound 11 configuration, and 8 features in the one trained on the rebound 7 labels. 7 features are present in both models: PL23, MS6\_80, MS0, PL24, MS2, MS6\_20, MS6\_200. Contributions from the features in the rebound 7 model are generally larger, also, the overall confidence of the model was higher. We noticed that the most impactful features were coming from the Market Shift features. 
       In Figure~\ref{fig:decPlotESH2019reb7} one can see two general decision patterns: one ending up at around 0.12-0.2 output probability and another one consisting of several misclassified entries ending up at around 0.8 output probability. In the first decision path, most of the MS features contributed consistently towards the negative class, however, PL23 and PL2 often pushed the probability in the opposite direction. In the second decision path, PL23 had the most persistent effect for the positive class, which was opposed by MS6\_80 in some cases and got almost no contribution from MS0 and MS6\_200 features. 
       
       Further analysis of ESH2019, rebound 11 configuration (omitted for brevity) showed that there was a skew in the output probabilities towards the negative class. Contributions from the PL features were less pronounced than for the rebound 7 model - the top 8 features belonged to the MS feature extraction step. There was no obvious decision path with misclassified entries. At the same time, we saw strong contributions toward negative outputs from MS6\_20 and MS6\_40. 
       
    \subsection{Simulated trading}
\begin{figure}[!htb]
            \centering
                    \includegraphics[width=0.8\textwidth]{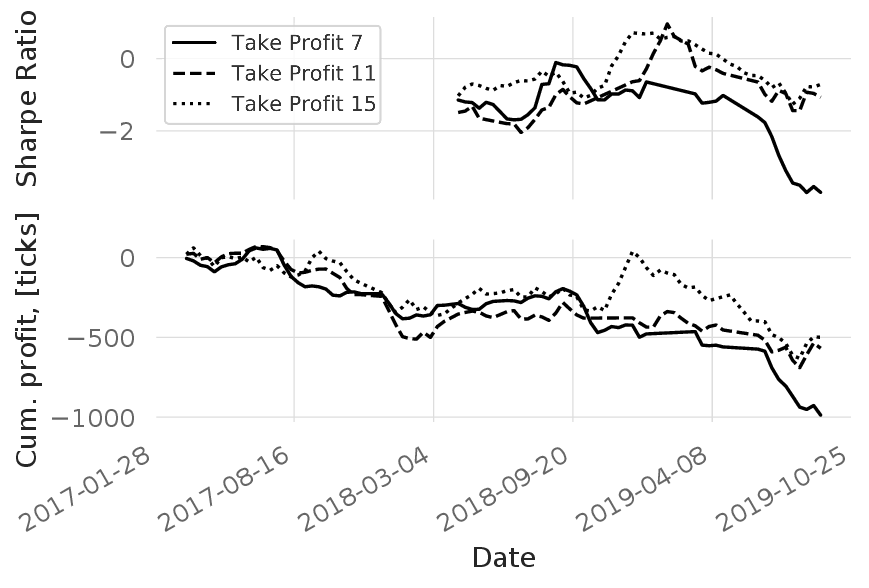}
                    \caption{Cumulative profit curves for the rebound of 15 ticks and take-profits of 7, 11 and 15 ticks for the years 2017-2019 with the corresponding annualised rolling Sharpe ratios (5\% risk-free income). The trading fees are already included.}
            \label{fig:profitSharpeyFixedRebound}
        \end{figure}

        In the current section, we report the cumulative profits for the 3 different labelling configurations in Fig.~\ref{fig:profitSharpeyFixedTP}. In addition to the cumulative profits, we report annualised Sharpe ratios with a 5\% risk-free annual profit. These results are reported for 15 ticks take profit, as shown in Fig.~\ref{fig:strata}. There was a descending trend in the Sharpe ratios across all the configurations. However, the behaviour of rebound 15 differed from the rest by performing worse until Oct 2018, then had a profitable period which was not observed for the other configurations.
        We also ran experiments with altered take profits for 15 ticks rebound labelling and found out that decreasing take profit led to slightly worse cumulative profits and comparable Sharpe ratios (Figure~\ref{fig:profitSharpeyFixedRebound}). It seems that the profitability period after Oct 2018 relied on the take profit value and vanished if the take profit is reduced.

        When computing the net outcomes of the trades, we add \$4.2 per-contract trading costs based on our assessment of current broker and clearance fees. 
        We did not set any slippage in the backtesting engine, since ES liquidity is large. However, we execute the stop-losses and take-profits on the tick following the close-position signal to account for order execution delays and slippages. This allowed for taking into account uncertainty rooting from large volatilities and gaps happening during the extraordinary market events. The backtesting was done on the tick data therefore there are no bar-backtesting assumptions made. 
        
\section{Discussion}
\label{sec:discussion}
This section breaks down and analyses the results presented in Sec.~\ref{sec:results}.
Results are discussed in relation to the overall pattern extraction and model performance, then research questions, model analysis and, finally, simulated trading.
Also, we discuss the obtained results in the context of the proposed methodology, as well as the limitations of every component and how they can be addressed. 
Finally, we present our view on implications for practitioners and our intuition of potential advancements and future work in this area.

    \subsection{Pattern extraction}
        The numbers of price levels and ticks per contract followed the same trend (Table~\ref{tab:processedEntries}). 
        Since the number of peaks was proportional to the number of ticks, we can say that the mean peak density is preserved over time to a large extent. In this context, the peak pattern can be considered stationary and appears in various market conditions.
    
    \subsection{Price Levels, CatBoost versus No-information estimator (RQ1)}
    \label{sec:Discussion:RQ1}
        The precision improvement for the CatBoost over the no-information estimator varied a lot across contracts (Table~\ref{tab:RQ12MSPprecs}). At the same time, the improvement was largely preserved across the labelling configurations. This might have been due to the original feature space, whose effectiveness relied a lot on the market state - in certain market states (and periods) the utility of the feature space dropped and the drop was consistent across the labelling configurations. Extensive research of the feature spaces would be necessary to make further claims. The overall performance of the models was weak on an absolute scale, however, it is comparable to the existing body of knowledge in the area of financial markets~\cite{Dixon2017Classification-basedNetworks}.
        
        Effect sizes in RQ1 Table~\ref{tab:RQ12testStats} were not significant. The significance here means that the chosen feature space with the model significantly contributed to the performance improvement with respect to the no-information model. A positive but insignificant effect size meant that there is an improvement that is limited to the considered sample and is unlikely to generalise to the unseen data (population).
        
        Assessing the results of the statistical tests, we used the significance level corrected for the multiple comparisons. Tests run on the rebound 7 and 15 configurations resulted in significant p-values, whilst rebound 11 was insignificant (Table~\ref{tab:RQ12testStats}). Hence, we rejected the null hypothesis H$_{01}$ for the labelling configurations of 7 and 15 ticks. This outcome was not supported by the effect sizes. 
        This divergence between the test and effect sizes indicated that the proposed trading pattern cannot be efficiently used in the machine learning setting with the considered feature space.
        Interestingly, standard deviations differ consistently between the compared groups across the configurations. While no-information model performance depends solely on the fraction of the positively labelled entries, CatBoost performance additionally depends on the suitability of the feature space and model parameters - this explains higher standard deviations in the CatBoost case.

    \subsection{Price Levels, 2-step feature extraction versus its components (RQ2)}
    \label{sec:Discussion:RQ2}
        For the effect sizes in Table~\ref{tab:RQ12testStats} we compared the 2-step feature extraction approach to its components - PL and MS. We did not see any significant effects showing supremacy of any of the approaches. Negative effect sizes in the case of the Market Shift component mean that in the considered sample MS performs better than the 2-step approach. This result did not generalise to the unseen data as its confidence intervals cross the 0-threshold. The possible explanation of the result is the much larger feature space of the 2-step approach (consisting of PL and MS features) than the MS compound. In case PL features were generally less useful than MS, which was empirically supported by our model analysis, this meant they might have had a negative impact during the feature selection process by introducing noise. 

        For assessing the statistical tests we used the significance level corrected for 6 comparisons. The p-values from Table~\ref{tab:RQ12testStats} show that there was no significant outcome and the null hypothesis H$_{02}$ could not be rejected. While the 2-step approach did not bring any improvement to the pipeline, there is no evidence that it significantly harmed the performance either. 
        It might have been the case that if PL features were designed differently, the method could have benefited from them. In this study, we withheld from iteratively tweaking the feature space to avoid any loss of statistical power. We find this aspect interesting for future work, however, it would require increasing the sample size to be able to account for the increased number of comparisons.
         
    \subsection{Simulated trading}
        In the simulated trading, we observed an interesting result - data labelling configuration had more impact on the profitability than the take profits (Figures~\ref{fig:profitSharpeyFixedRebound},\ref{fig:profitSharpeyFixedTP}). We hypothesise that the reason for this is that the simplistic trading strategy is overused by the trading community in various configurations. 
        In contrast, the labelling configuration is less straightforward and had more impact on profitability.
        Note that the obtained precision (Table~\ref{tab:RQ12MSPprecs}) could not be directly related to the considered trading strategy as there might be multiple price levels extracted within the time interval of a single trade. Consequently, there might have been extrema which were not traded. 
        
        It is hard to expect consistent profitability considering the simplicity of the strategy and lack of optimisation of the feature space, however, even in the current setting one can see profitable episodes (Fig~\ref{fig:profitSharpeyFixedTP}). 
        The objective of the study was not to provide a ready-to-trade strategy, but rather to showcase the proposed methodology.
        We believe that the demonstrated approach is generalisable to other trading strategies.
     
     \subsection{Methodology}
        The point we have highlighted in this study is that by looking at the machine learning model performance or simulated trading results on their own, one might be drawn into believing that the considered trading pattern and the feature extraction method might work. 
        This, consequently, would lead to additional time, effort and financial risks until one discovers that the applications of the proposed pattern are limited. 
        
        On the other hand, using the proposed methodology, one could withhold further development at the point of computing the effect sizes and the confidence intervals. Not only does the proposed methodology optimises the whole automated trading platform component design approach, but it also allows uniform and comparable reporting of the component design.
        
        As it was shown in Section~\ref{sec:backMethodATP}, profitability is often used as a measure of success when evaluating trading patterns and trading strategies. 
        The profitability of the price levels pattern might look promising for certain intervals of the considered timeline (Figure~\ref{fig:profitSharpeyFixedRebound}), especially considering that the trading fees are taken into account. However, being guided by the proposed methodology, it becomes clear that the considered pattern does not warrant attention in the considered setting. 
        This is a common example of how the profitability metric itself might be misleading if reported on its own.
        
    \subsection{Limitations}
    \label{sec:Discussion:Limitations}
        The proposed methodology is one of the many ways the financial markets can be studied empirically. We encourage practitioners to adjust our proposed methodology to their needs and setting while keeping the interpretability and comparability of the study in mind. 
        
        Depending on the data distributions, size of the data sample, etc., one might use different statistical tests and effect size measures without any loss of interpretability.
        For example, while it is advised to use Glass's $\Delta$ in case of the significantly different standard deviations between groups, this measure does not have corrections for the paired data. Hence, in our experiment design, we choose to stick to the Hedge's \textit{g}$_{av}$. 
        For the sake of completeness, we verified the results using Glass's $\Delta$ - and 15 ticks rebound effect size becomes significant in RQ1.
    
        In the backtesting we used the last trade price to define ticks, we do not take into account bid-ask spreads. In live trading, trades as executed by bid or ask price, depending on the direction of the trade. It leads to a hidden fee of the bid-ask spread per trade. This is crucial for intraday trading as average profits per trade often lay within a couple of ticks.
        Moreover, when modelling order executions, we do not consider per-tick volumes coming from aggressive buyers and sellers (bid and ask). It might be the case that for some ticks only aggressive buyers (sellers) were present, and our algorithm executed a long (short) limit order.
        This leads to uncertainty in opening positions - in reality, some of the profitable orders may have not been filled. At the same time, losing orders would always be executed. 

        Another limitation is that we do not model order queues, and, consequently cannot guarantee that our orders would have been filled if we submitted them live even if both bid and ask volumes were present in the tick.  
        This is crucial for high-frequency trading (HFT), where thousands of trades are performed daily with tiny take-profits and stop-losses, but has less impact on the trade intervals considered in the study. 
        Finally, there is an assumption that our entering the market does not change its state significantly. 
        We believe it is a valid assumption considering the liquidity of S\&P E-mini futures.
    
    \subsection{Implications for practitioners}
        We provide a systematic approach to the evaluation of automated trading platform components. While large market participants have internal evaluation procedures, we believe that our research could support various existing pipelines. Considering the state of the matter with the lack of code and data publishing in the field, we are confident that the demonstrated approach can be used towards improving the generalisability and reproducibility of research. Specific methods like extrema classification and 2-step feature extraction can serve as a baseline for the effect sizes observed in the field. 
            
    \subsection{Future work}
        While the proposed methodology is in an initial stage for the field of financial markets, we believe that it is flexible enough to be incorporated into a python package for trading pattern design. That would significantly boost its adoption by practitioners. 
    
        In the current study, we  evaluate the utility of the price levels trading pattern.
        The next step would be to propose a more holistic pattern. 
        We would aim to extend the market properties to volumes and volume profiles.
        
        Another direction is to assess the price levels for trading trends (instead of reversals) - in this case, one would aim to classify price level crossings with a statistically significant improvement with respect to the no-information model. A different definition of the price crossing at the point of data labelling is necessary for that. 
    
        In terms of improving the strategy, there is a couple of things that can be done.
        For instance: take-profit and stop-loss offsets might be linked to the volatility instead of being constant. 
        Also, flat strategies usually work better at certain times of the day - it would be wise to interrupt trading before USA and EU session starts and ends, as well as scheduled reports, news and impactful speeches. 
        Additionally, all the mentioned parameters we have chosen can be looked into and optimised to the needs of the market participant.   
        
        It would be interesting to investigate other means for feature design instead of the manually defined feature space. For instance, Deep Learning networks can be used to obtain market representations at every moment of time~\cite{wang2020stock2vec}. It should be noted though that such an approach would substantially limit the transparency of the pipeline. 
        
        In terms of the chosen model, it would be useful comparing the CatBoost classifier to deep learning models like DA-RNN~\cite{Qin2017}, as it makes use of the attention-based architecture designed based on the recent breakthrough in the area of natural language processing \cite{radford2019language}. Moreover, it would be useful to consider model calibration approaches to obtain more realistic output class probabilities from the model~\cite{vaicenavicius2019evaluating}.
        
        Finally, we see a gap in the available FLOSS (Free/Libre Open Source Software) backtesting tools. To the best of our knowledge, there is no publicly available backtesting engine taking into account bid and ask prices and order queues. While there are solutions with this functionality provided as parts of the proprietary trading platforms, they can only be used as a black box. An open-source engine would contribute to transparency and has the potential to become the solution for both the research and industry worlds.
        
\section{Conclusion}

Our work proposed a generic methodology for the systematic evaluation of APT components. Then, it showcased the proposed approach to the trading pattern and the feature space design ATP components.

Whilst extrema have been discussed as potentially high-performance means for trading decisions, there has been no work proposing their automatic extraction and assessment from the statistical point of view. 
While the pattern is commonly used by practitioners in various settings and might be indeed useful for manual trading, its applications in the machine learning setting are only partially supported by statistics.
While the proposed 2-step feature extraction method is inspired by domain knowledge and market understanding, there is no statistical evidence that it benefits ATP.  

This paper has presented every single aspect of data processing, feature extraction, feature evaluation and selection, machine learning estimator optimisation and training, as well as details of the trading strategy.
Moreover, we statistically assessed the findings. We rejected the null hypothesis answering RQ1 - our approach performs statistically better than the baseline, however, the effect sizes stay non-significant. 
We did not observe any significant effect sizes for RQ2 and could not reject the null hypothesis. Hence, the use of the 2-step feature extraction does not improve the performance of the approach for the proposed feature space and the model. There is no evidence that it hurts the performance either. 
We hope that the proposed methodology and the showcased scenarios will enable the adoption of a more uniform and sound methodology in the field of financial markets.
We conclude by providing samples of our code online~\cite{zenodo2020paper}.

\section*{Acknowledgement}
The research is supported by CRITiCaL - Combatting cRiminals In The CLoud, under grant EP/M020576/1 and by AISEC: AI Secure and Explainable by Construction under grant  EP/T027037/1.
The authors would like to thank Roberto Metere for his drawing contributions.

\section*{Disclosures}
\textbf{Conflict of Interest:} The authors declare that they have no known competing financial interests or personal relationships that could have appeared to influence the work reported in this paper.

\bibliographystyle{plain}
\bibliography{bibliography}

\newpage \section*{Supplementary Materials}
\label{sec:supplementary}
\setcounter{table}{0}
\renewcommand\thetable{S\arabic{table}}
\setcounter{figure}{0}
\renewcommand\thefigure{S\arabic{figure}}

\begin{table}[!htb]
    \centering
    \footnotesize
    \caption{Model performance measures reported for the 2-step feature extraction, 3 different labelling configurations: 7, 11 and 15 ticks rebounds. Null-Precision corresponds to the performance of an always-positive classifier.}
    \begin{tabular}{l|rrrrr}

Contract &  PR-AUC &  F1-score &  Precision &  ROC-AUC &  Null-Precision \\
\hline
\multicolumn{6}{c}{\textbf{Rebound 7}} \\
\hline
ESH2017 &    .19 &      .26 &       .20 &     .51 &            .19 \\
ESM2017 &    .25 &      .28 &       .22 &     .54 &            .21 \\
ESU2017 &    .25 &      .29 &       .25 &     .56 &            .20 \\
ESZ2017 &    .18 &      .16 &       .17 &     .52 &            .16 \\
ESH2018 &    .19 &      .28 &       .19 &     .54 &            .17 \\
ESM2018 &    .22 &      .32 &       .21 &     .55 &            .19 \\
ESU2018 &    .17 &      .18 &       .17 &     .52 &            .16 \\
ESZ2018 &    .18 &      .28 &       .18 &     .53 &            .17 \\
ESH2019 &    .18 &      .24 &       .18 &     .51 &            .18 \\
ESM2019 &    .17 &      .24 &       .18 &     .52 &            .17 \\
\hline
\multicolumn{6}{c}{\textbf{Rebound 11}} \\
\hline
ESH2017 &    .19 &      .30 &       .20 &     .53 &            .18 \\
ESM2017 &    .21 &      .14 &       .23 &     .54 &            .19 \\
ESU2017 &    .18 &      .14 &       .15 &     .50 &            .19 \\
ESZ2017 &    .17 &      .23 &       .17 &     .52 &            .16 \\
ESH2018 &    .19 &      .27 &       .18 &     .53 &            .17 \\
ESM2018 &    .21 &      .31 &       .21 &     .55 &            .19 \\
ESU2018 &    .16 &      .17 &       .16 &     .50 &            .16 \\
ESZ2018 &    .19 &      .27 &       .17 &     .52 &            .17 \\
ESH2019 &    .18 &      .24 &       .18 &     .52 &            .17 \\
ESM2019 &    .17 &      .25 &       .17 &     .52 &            .17 \\
\hline
\multicolumn{6}{c}{\textbf{Rebound 15}} \\
\hline
ESH2017 &    .15 &      .20 &       .15 &     .51 &            .15 \\
ESM2017 &    .18 &      .21 &       .19 &     .53 &            .17 \\
ESU2017 &    .17 &      .17 &       .18 &     .54 &            .15 \\
ESZ2017 &    .17 &      .11 &       .19 &     .51 &            .16 \\
ESH2018 &    .17 &      .26 &       .17 &     .52 &            .16 \\
ESM2018 &    .18 &      .28 &       .18 &     .53 &            .17 \\
ESU2018 &    .16 &      .22 &       .16 &     .51 &            .16 \\
ESZ2018 &    .17 &      .26 &       .16 &     .52 &            .16 \\
ESH2019 &    .17 &      .26 &       .18 &     .51 &            .17 \\
ESM2019 &    .17 &      .26 &       .17 &     .52 &            .16 \\

\end{tabular}
    \label{tab:PrecsSuppl}
\end{table}

\begin{table}[!htb]
    \centering
    \footnotesize
    \caption{Model performance measures reported for the Price Level feature extraction component, 3 different labelling configurations: 7, 11 and 15 ticks rebounds. Null-Precision corresponds to the performance of an always-positive classifier.}
    \begin{tabular}{l|rrrrr}

Contract &  PR-AUC &  F1-score &  Precision &  ROC-AUC &  Null-Precision \\
\hline
\multicolumn{6}{c}{\textbf{Rebound 7}} \\
\hline
ESH2017 &    .19 &      .24 &       .19 &     .50 &            .19 \\
ESM2017 &    .21 &      .28 &       .21 &     .50 &            .21 \\
ESU2017 &    .22 &      .25 &       .23 &     .53 &            .20 \\
ESZ2017 &    .16 &      .19 &       .16 &     .49 &            .16 \\
ESH2018 &    .18 &      .23 &       .17 &     .50 &            .17 \\
ESM2018 &    .19 &      .25 &       .19 &     .49 &            .19 \\
ESU2018 &    .16 &      .15 &       .15 &     .50 &            .16 \\
ESZ2018 &    .18 &      .24 &       .17 &     .51 &            .17 \\
ESH2019 &    .18 &      .26 &       .18 &     .51 &            .18 \\
ESM2019 &    .18 &      .21 &       .18 &     .50 &            .17 \\
\hline
\multicolumn{6}{c}{\textbf{Rebound 11}} \\
\hline
ESH2017 &    .17 &      .19 &       .17 &     .49 &            .18 \\
ESM2017 &    .20 &      .18 &       .19 &     .51 &            .19 \\
ESU2017 &    .18 &      .18 &       .17 &     .50 &            .19 \\
ESZ2017 &    .16 &      .19 &       .16 &     .50 &            .16 \\
ESH2018 &    .17 &      .23 &       .18 &     .51 &            .17 \\
ESM2018 &    .19 &      .17 &       .20 &     .50 &            .19 \\
ESU2018 &    .16 &      .22 &       .16 &     .50 &            .16 \\
ESZ2018 &    .18 &      .20 &       .18 &     .52 &            .17 \\
ESH2019 &    .18 &      .25 &       .17 &     .51 &            .17 \\
ESM2019 &    .17 &      .22 &       .17 &     .50 &            .17 \\
\hline
\multicolumn{6}{c}{\textbf{Rebound 15}} \\
\hline
ESH2017 &    .15 &      .20 &       .15 &     .50 &            .15 \\
ESM2017 &    .17 &      .24 &       .18 &     .51 &            .17 \\
ESU2017 &    .17 &      .16 &       .19 &     .53 &            .15 \\
ESZ2017 &    .16 &      .13 &       .15 &     .50 &            .16 \\
ESH2018 &    .16 &      .25 &       .16 &     .50 &            .16 \\
ESM2018 &    .18 &      .21 &       .17 &     .51 &            .17 \\
ESU2018 &    .15 &      .16 &       .15 &     .49 &            .16 \\
ESZ2018 &    .16 &      .19 &       .16 &     .51 &            .16 \\
ESH2019 &    .17 &      .24 &       .17 &     .50 &            .17 \\
ESM2019 &    .16 &      .18 &       .16 &     .49 &            .16 \\

\end{tabular}
    \label{tab:AllMetricsPSuppl}
\end{table}

\begin{table}[!htb]
    \centering
    \footnotesize
    \caption{Model performance measures reported for the Market Shift feature extraction component, 3 different labelling configurations: 7, 11 and 15 ticks rebounds. Null-Precision corresponds to the performance of an always-positive classifier.}
    \begin{tabular}{l|rrrrr}

Contract &  PR-AUC &  F1-score &  Precision &  ROC-AUC &  Null-Precision \\
\hline
\multicolumn{6}{c}{\textbf{Rebound 7}} \\
\hline
ESH2017 &    .20 &      .26 &       .20 &     .51 &            .19 \\
ESM2017 &    .25 &      .28 &       .24 &     .55 &            .21 \\
ESU2017 &    .25 &      .31 &       .25 &     .56 &            .20 \\
ESZ2017 &    .17 &      .26 &       .18 &     .52 &            .16 \\
ESH2018 &    .20 &      .29 &       .19 &     .54 &            .17 \\
ESM2018 &    .21 &      .33 &       .22 &     .55 &            .19 \\
ESU2018 &    .17 &      .23 &       .17 &     .51 &            .16 \\
ESZ2018 &    .18 &      .28 &       .18 &     .52 &            .17 \\
ESH2019 &    .19 &      .28 &       .19 &     .53 &            .18 \\
ESM2019 &    .18 &      .27 &       .18 &     .53 &            .17 \\
\hline
\multicolumn{6}{c}{\textbf{Rebound 11}} \\
\hline
ESH2017 &    .18 &      .24 &       .19 &     .51 &            .18 \\
ESM2017 &    .21 &      .24 &       .20 &     .50 &            .19 \\
ESU2017 &    .21 &      .31 &       .23 &     .56 &            .19 \\
ESZ2017 &    .16 &      .12 &       .16 &     .50 &            .16 \\
ESH2018 &    .20 &      .28 &       .18 &     .54 &            .17 \\
ESM2018 &    .20 &      .32 &       .21 &     .55 &            .19 \\
ESU2018 &    .17 &      .22 &       .17 &     .51 &            .16 \\
ESZ2018 &    .17 &      .27 &       .17 &     .52 &            .17 \\
ESH2019 &    .18 &      .26 &       .18 &     .52 &            .17 \\
ESM2019 &    .18 &      .27 &       .18 &     .53 &            .17 \\
\hline
\multicolumn{6}{c}{\textbf{Rebound 15}} \\
\hline
ESH2017 &    .15 &      .22 &       .15 &     .48 &            .15 \\
ESM2017 &    .17 &      .20 &       .16 &     .50 &            .17 \\
ESU2017 &    .16 &      .17 &       .15 &     .51 &            .15 \\
ESZ2017 &    .16 &      .11 &       .16 &     .50 &            .16 \\
ESH2018 &    .17 &      .27 &       .17 &     .53 &            .16 \\
ESM2018 &    .18 &      .28 &       .18 &     .54 &            .17 \\
ESU2018 &    .16 &      .22 &       .16 &     .51 &            .16 \\
ESZ2018 &    .17 &      .27 &       .17 &     .52 &            .16 \\
ESH2019 &    .17 &      .27 &       .17 &     .52 &            .17 \\
ESM2019 &    .18 &      .27 &       .18 &     .53 &            .16 \\

\end{tabular}
    \label{tab:AllMetricsMSSuppl}
\end{table}

\begin{figure}[!htb]
\centering
   \begin{minipage}{0.8\textwidth}
     \centering
    \includegraphics[width=0.9\textwidth]{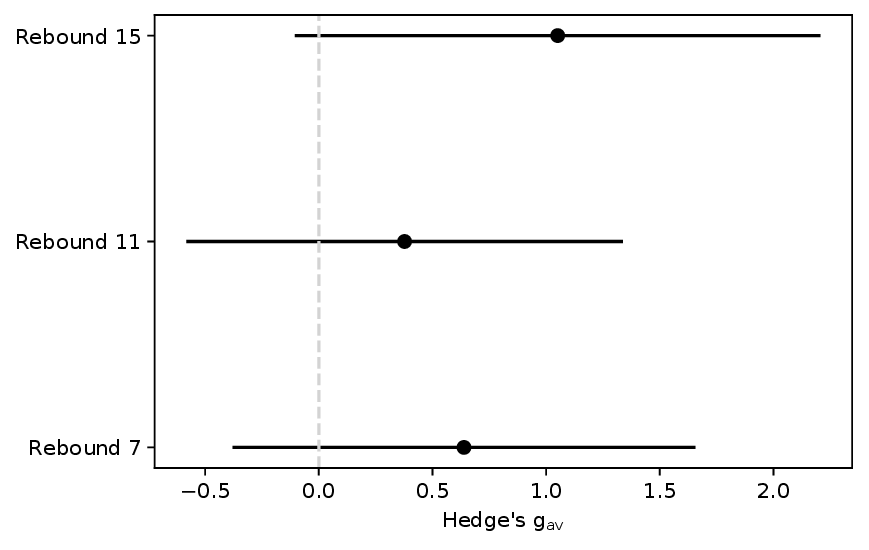}
    \caption{Hedge's g$_{av}$ effect sizes quantify the improvement of the precision from using the CatBoost over the no-information estimator. The error bars illustrate the .95 confidence intervals, corrected for multiple comparisons (3 in this case). The dashed line corresponds to the significance threshold. Rebounds accord to different labelling configurations, where 7, 11 and 15 are ticks required for the positive labelling of an entry.}
   \end{minipage}\hfill
   \begin{minipage}{0.8\textwidth}
     \centering
    \includegraphics[width=0.9\textwidth]{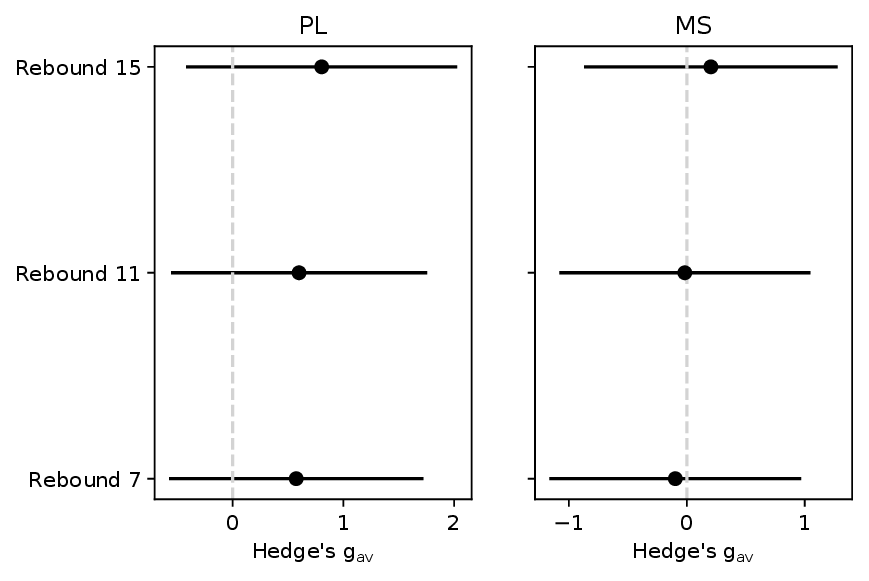}
    \caption{Hedge's g$_{av}$ effect sizes quantify the improvement of the precision from using the 2-step feature extraction over each of the components (PL and MS). The error bars illustrate the .95 confidence intervals, corrected for multiple comparisons (6 in this case). The dashed line corresponds to the significance threshold. Rebounds accord to different labelling configurations, where 7, 11 and 15 are ticks required for the positive labelling of an entry.}
    \label{fig:RQ2EffSize}
   \end{minipage}
\end{figure}

\begin{figure}[!htb]
\centering
   \begin{minipage}{0.48\textwidth}
     \centering
    \includegraphics[width=\textwidth]{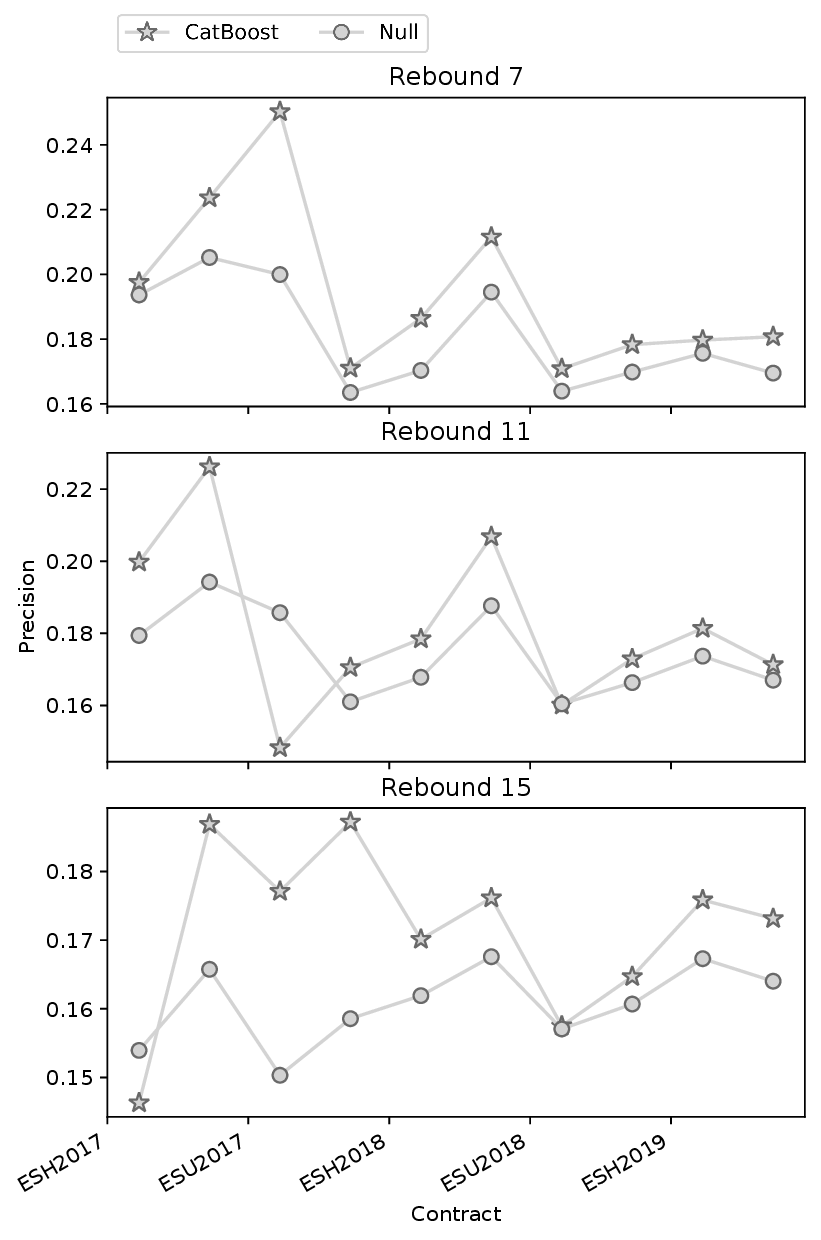}
  \caption{The precision of the CatBoost model and the always-positive estimator. The plot shows the labelling configurations of 7, 11 and 15 ticks rebounds.}
    \label{fig:RQ1PrecSuppl}
   \end{minipage}\hfill
   \begin{minipage}{0.48\textwidth}
     \centering
    \includegraphics[width=\textwidth]{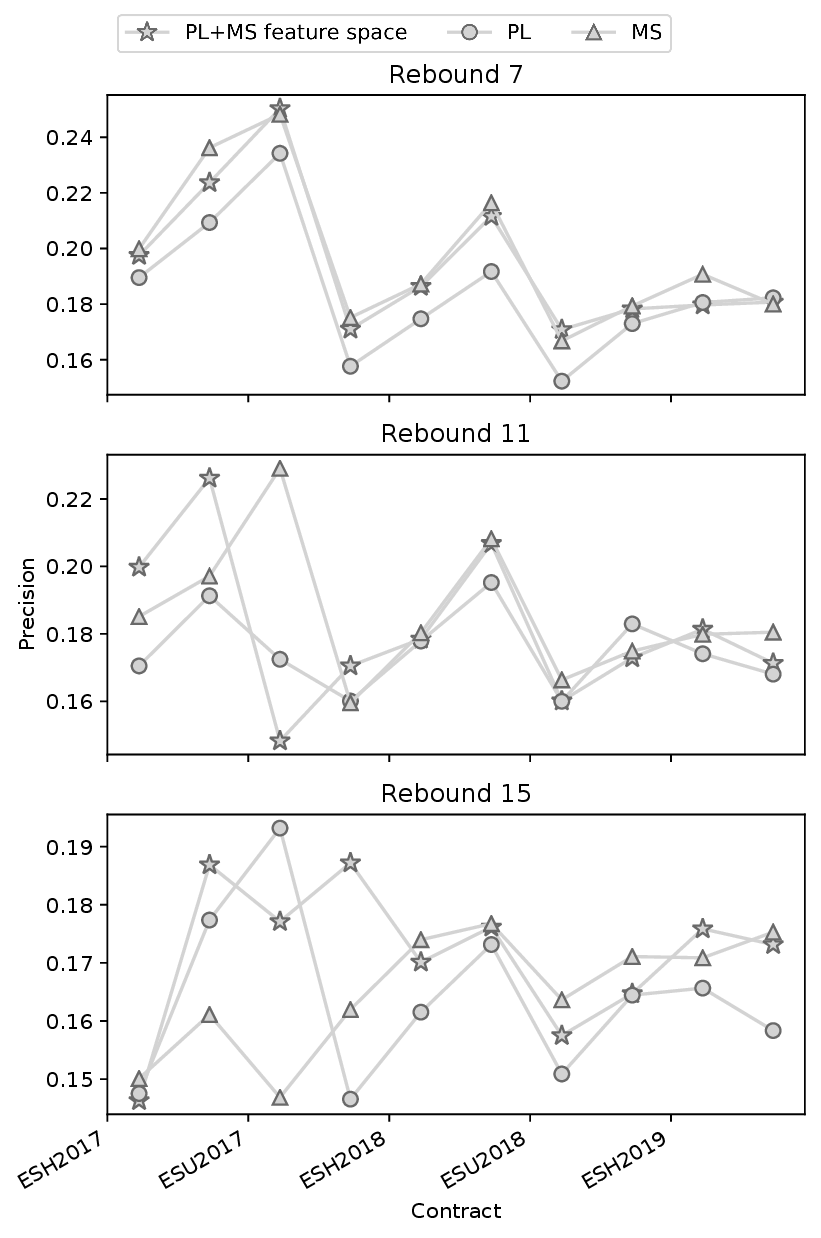}
    \caption{The precision of the model which uses the 2-step feature extraction (MS+PL) versus the performance of the models using the single-step feature extraction (MS, PL). The plot shows the labelling configurations of 7, 11 and 15 ticks rebounds.}
    \label{fig:RQ2PrecSuppl}
   \end{minipage}
\end{figure}

\begin{figure}[!htb]
    \centering
    \includegraphics[width=0.75\textwidth]{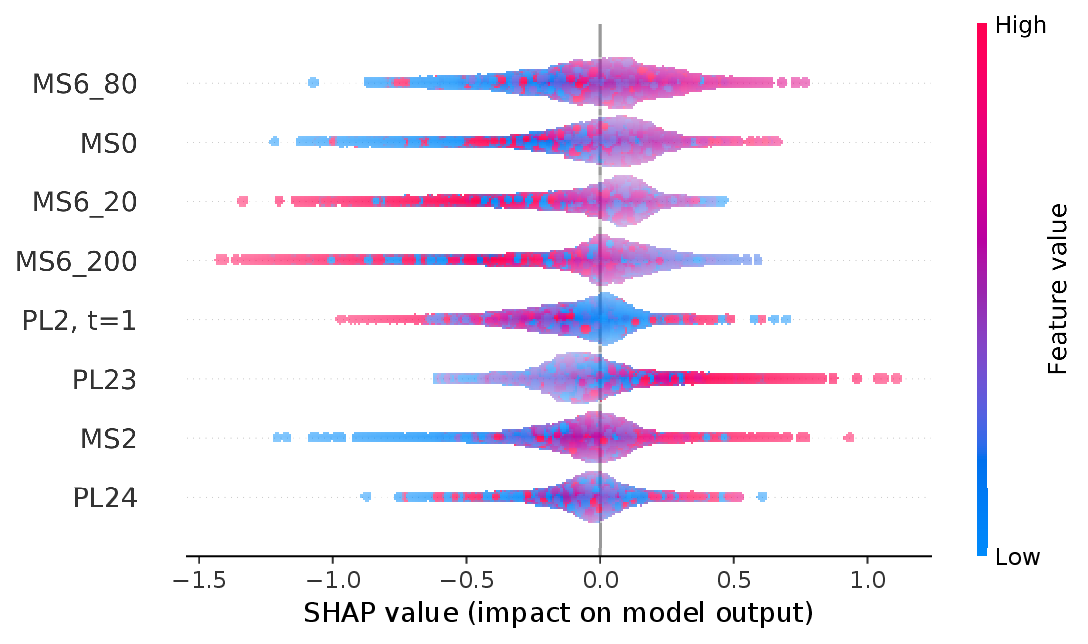}
    \caption{SHAP summary plot of the model trained on ESH2019 contract, rebound 7 configuration. Each marker is a classified entry. The X-axis quantifies the contribution of the entries towards the positive or negative class output.}
    \label{fig:SHAPsummayESH2019reb7Suppl}
\end{figure}

\begin{figure}[!htb]
    \centering
            \includegraphics[width=0.75\textwidth]{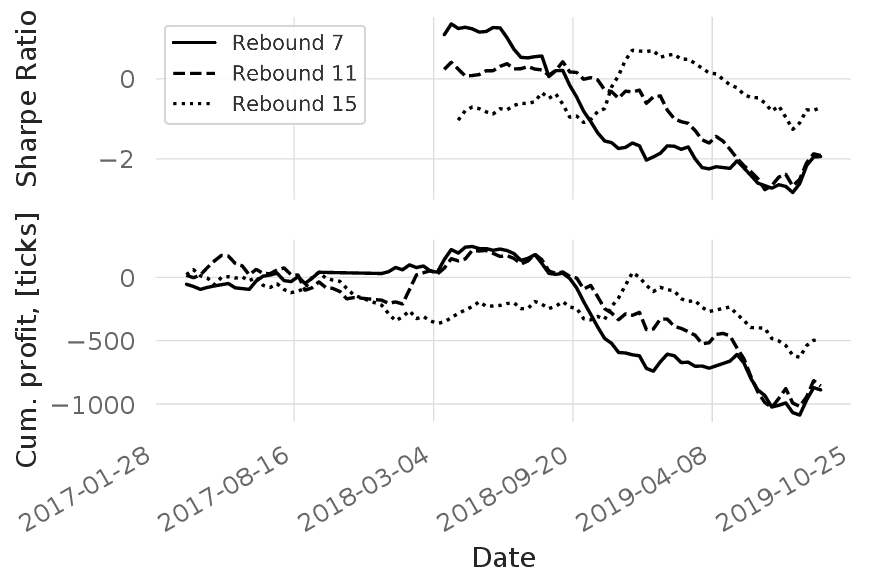}
            
            \caption{Cumulative profit curves for all the rebound configurations and fixed take profit of 15 ticks for years 2017-2019 with the corresponding annualised rolling Sharpe ratios (computed for 5\% risk-free income). The trading fees are already included in the cumulative profits.}
    \label{fig:profitSharpeyFixedTP}
\end{figure}

\end{document}